\documentclass[10pt,aps,prd,twocolumn,a4paper,showkeys,nofootinbib]{revtex4-1}

\usepackage[T1]{fontenc}
\usepackage{times}
\DeclareMathAlphabet{\MATHIT}{OT1}{ptm}{m}{it}
\DeclareSymbolFont{Letters}{OML}{ztmcm}{m}{it}
\DeclareSymbolFontAlphabet{\mathNormal}{Letters}
\usepackage[utf8]{inputenc}
\usepackage[ngerman,english]{babel}
\usepackage{amsmath}
\usepackage{amsfonts}
\usepackage{amssymb}
\usepackage{amsthm}
\usepackage{enumitem}
\usepackage{dcolumn}
\usepackage{bm}
\usepackage{bbm}
\usepackage{cancel}
\usepackage{tikz}
\usetikzlibrary{calc,shapes,arrows,patterns,decorations.pathmorphing,positioning,automata}
\usepackage{tabularx}
\usepackage{booktabs}
\usepackage{tensor}
\usepackage{scalerel}
\usepackage{xcolor}
\usepackage{mathtools}
\usepackage{float}
\usepackage{xfrac}
\usepackage[pdfborder={0 0 0}]{hyperref}
\definecolor{darkblue}{rgb}{0,0,.5}
\definecolor{darkgreen}{rgb}{0,0.5,.5}
\definecolor{darkyellow}{rgb}{0.5,0.5,0}
\definecolor{fhl}{rgb}{1,0,0}
\hypersetup{colorlinks=true, breaklinks=true, linkcolor=darkblue, menucolor=darkblue, urlcolor=darkblue, linktocpage=true}
\usepackage{geometry}
\usepackage{textpos}
\usepackage{relsize}
\usepackage{balance}
\usepackage[titletoc,title]{appendix}
\allowdisplaybreaks
\makeatletter 
\newsavebox\myboxA 
\newsavebox\myboxB 
\newlength\mylenA 
 
\newcommand*\xoverline[2][0.75]{%
    \sbox{\myboxA}{$\m@th#2$}%
    \setbox\myboxB\null
    \ht\myboxB=\ht\myboxA%
    \dp\myboxB=\dp\myboxA%
    \wd\myboxB=#1\wd\myboxA
    \sbox\myboxB{$\m@th\overline{\copy\myboxB}$}
    \setlength\mylenA{\the\wd\myboxA}
    \addtolength\mylenA{-\the\wd\myboxB}%
    \ifdim\wd\myboxB<\wd\myboxA%
       \rlap{\hskip 0.5\mylenA\usebox\myboxB}{\usebox\myboxA}%
    \else 
        \hskip -0.5\mylenA\rlap{\usebox\myboxA}{\hskip 0.5\mylenA\usebox\myboxB}%
    \fi}
\makeatother
\numberwithin{equation}{section}

\let\originalleft\left
\let\originalright\right
\renewcommand{\left}{\mathopen{}\mathclose\bgroup\originalleft}
\renewcommand{\right}{\aftergroup\egroup\originalright}
\newcommand{\e}{\operatorname{e}}

\newcommand{\SU}[1]{\operatorname{SU}(#1)}

\newcommand{\On}[1]{\operatorname{O}\left(#1\right)}
\newcommand{\Un}[1]{\operatorname{U}\left(#1\right)}

\newcommand{\CP}[1]{\operatorname{CP}\left(#1\right)}

\newcommand{\Sph}[1]{\operatorname{S}^{#1}}
\newcommand{\of}[1]{\mathchoice{\left(#1\right)}{(#1)}{(#1)}{(#1)}}

\newcommand{\loint}[1]{\left(#1\right]}
\newcommand{\roint}[1]{\left[#1\right)}

\newcommand{\bof}[1]{\biggl(\bigg.#1\bigg.\biggr)}

\newcommand{\sof}[1]{\bigl(\big.#1\big.\bigr)}
\newcommand{\ssof}[1]{(#1)}
\newcommand{\fof}[1]{\left[#1\right]}

\newcommand{\ssfof}[1]{[#1]}
\newcommand{\cof}[1]{\left\{#1\right\}}

\newcommand{\bcof}[1]{\biggl\{\bigg.#1\bigg.\biggr\}}
\newcommand{\sbcof}[1]{\Bigl\{\Big.#1\Big.\Bigr\}}

\newcommand{\sscof}[1]{\{#1\}}

\newcommand{\avof}[1]{\left\langle #1\right\rangle}

\newcommand{\savof}[1]{\big\langle #1\big\rangle}
\newcommand{\ssavof}[1]{\langle \smash{#1}\rangle}

\DeclarePairedDelimiter\floor{\lfloor}{\rfloor}
\newcommand{\diag}{\operatorname{diag}}
\newcommand{\trace}{\operatorname{tr}}

\newcommand{\ii}{\mathrm{i}}
\newcommand{\idd}[2]{\mathrm{d}^{#2}#1}
\newcommand{\dd}{\mathrm{d}}
\newcommand{\DD}[1]{\mathcal{D}\bigl[#1\bigr]}

\newcommand{\totd}[2]{\frac{\dd #1}{\dd #2}}

\newcommand{\partd}[2]{\frac{\partial #1}{\partial #2}}

\newcommand{\partdm}[3]{\frac{\partial^{#3} #1}{\of{\partial #2}^{#3}}}

\newcommand{\spartd}[3]{\frac{\partial^{2} #1}{\partial #2 \partial #3}}

\newcommand{\order}[1]{\mathcal{O}(#1)}

\newcommand{\sech}{\operatorname{sech}}

\newcommand{\sgn}{\operatorname{sgn}}

\newcommand{\abs}[1]{\left| #1\right|}
\newcommand{\ssabs}[1]{| #1|}
\newcommand{\sabs}[1]{\big| #1\big|}

\newcommand{\mbeq}{\overset{!}{=}}

\renewcommand*\[{\begin{equation}}
\renewcommand*\]{\end{equation}}
\renewcommand*\hat[1]{\smash{\widehat{#1}}}
\let\oldstackrel\stackrel
\renewcommand*\stackrel[2]{{\scriptstyle\oldstackrel{#1}{#2}}}

\definecolor{emphcol}{RGB}{0,0,0}
\let\oldemph\emph
\renewcommand*\emph[1]{\oldemph{\textcolor{emphcol}{#1}}}
\let\oldstackrel\stackrel
\renewcommand*\stackrel[2]{{\scriptstyle\oldstackrel{#1}{#2}}}

\newcommand{\ucases}[1]{\begin{cases}#1\end{cases}}
\renewcommand*\doi[1]{DOI:~\href{https://doi.org/#1}{#1}}
\newcommand{\arxiv}[1]{arXiv:~\href{https://arxiv.org/abs/#1}{#1}}

\newlength{\hatchspread}
\newlength{\hatchthickness}
\newlength{\hatchshift}
\newcommand{\hatchcolor}{}
\tikzset{hatchspread/.code={\setlength{\hatchspread}{#1}},
         hatchthickness/.code={\setlength{\hatchthickness}{#1}},
         hatchshift/.code={\setlength{\hatchshift}{#1}},
         hatchcolor/.code={\renewcommand{\hatchcolor}{#1}}}
\tikzset{hatchspread=3pt,
         hatchthickness=0.4pt,
         hatchshift=0pt,
         hatchcolor=black}
\pgfdeclarepatternformonly[\hatchspread,\hatchthickness,\hatchshift,\hatchcolor]
   {custom north west lines}
   {\pgfqpoint{\dimexpr-2\hatchthickness}{\dimexpr-2\hatchthickness}}
   {\pgfqpoint{\dimexpr\hatchspread+2\hatchthickness}{\dimexpr\hatchspread+2\hatchthickness}}
   {\pgfqpoint{\dimexpr\hatchspread}{\dimexpr\hatchspread}}
   {
    \pgfsetlinewidth{\hatchthickness}
    \pgfpathmoveto{\pgfqpoint{\dimexpr0pt}{\dimexpr\hatchspread+\hatchshift}}
    \pgfpathlineto{\pgfqpoint{\dimexpr\hatchspread+0.15pt}{\dimexpr\hatchshift-0.15pt}}
    \ifdim \hatchshift > 0pt
      \pgfpathmoveto{\pgfqpoint{\dimexpr-0.15pt}{\dimexpr\hatchshift+0.15pt}}
      \pgfpathlineto{\pgfqpoint{\dimexpr\hatchshift+0.15pt}{\dimexpr-0.15pt}}
    \fi
    \pgfsetstrokecolor{\hatchcolor}
    \pgfusepath{stroke}
   }

\pgfdeclarepatternformonly[\hatchspread,\hatchthickness,\hatchshift,\hatchcolor]
   {custom north east lines}
   {\pgfqpoint{\dimexpr-2\hatchthickness}{\dimexpr-2\hatchthickness}}
   {\pgfqpoint{\dimexpr\hatchspread+2\hatchthickness}{\dimexpr\hatchspread+2\hatchthickness}}
   {\pgfqpoint{\dimexpr\hatchspread}{\dimexpr\hatchspread}}
   {
    \pgfsetlinewidth{\hatchthickness}
    \pgfpathmoveto{\pgfqpoint{0pt}{\dimexpr\hatchshift}}
    \pgfpathlineto{\pgfqpoint{\dimexpr\hatchspread+0.15pt}{\dimexpr\hatchspread+0.15pt+\hatchshift}}
    \ifdim \hatchshift > 0pt
      \pgfpathmoveto{\pgfqpoint{\dimexpr\hatchspread-\hatchshift-0.15pt}{\dimexpr-0.15pt}}
      \pgfpathlineto{\pgfqpoint{\dimexpr\hatchspread+0.15pt}{\dimexpr\hatchshift+0.15pt}}
    \fi
    \pgfsetstrokecolor{\hatchcolor}
    \pgfusepath{stroke}
   }

\pgfdeclarepatternformonly[\hatchspread,\hatchthickness,\hatchshift,\hatchcolor]
   {custom vertical lines}
   {\pgfqpoint{\dimexpr-2\hatchthickness}{\dimexpr-2\hatchthickness}}
   {\pgfqpoint{\dimexpr\hatchspread+2\hatchthickness}{\dimexpr\hatchspread+2\hatchthickness}}
   {\pgfqpoint{\dimexpr\hatchspread}{\dimexpr\hatchspread}}
   {
    \pgfsetlinewidth{\hatchthickness}
    \pgfpathmoveto{\pgfqpoint{\dimexpr\hatchshift}{0pt}}
    \pgfpathlineto{\pgfqpoint{\dimexpr\hatchshift}{\dimexpr\hatchspread+0.15pt}}
    \pgfsetstrokecolor{\hatchcolor}
    \pgfusepath{stroke}
   }

\pgfdeclarepatternformonly[\hatchspread,\hatchthickness,\hatchshift,\hatchcolor]
   {custom horizontal lines}
   {\pgfqpoint{\dimexpr-2\hatchthickness}{\dimexpr-2\hatchthickness}}
   {\pgfqpoint{\dimexpr\hatchspread+2\hatchthickness}{\dimexpr\hatchspread+2\hatchthickness}}
   {\pgfqpoint{\dimexpr\hatchspread}{\dimexpr\hatchspread}}
   {
    \pgfsetlinewidth{\hatchthickness}
    \pgfpathmoveto{\pgfqpoint{0pt}{\dimexpr\hatchshift}}
    \pgfpathlineto{\pgfqpoint{\dimexpr\hatchspread+0.15pt}{\dimexpr\hatchshift}}
    \pgfsetstrokecolor{\hatchcolor}
    \pgfusepath{stroke}
   }
 
\tikzset{cross/.style={cross out,draw,minimum size=2*(#1-\pgflinewidth),inner sep=0pt, outer sep=0pt}}

\begin{document}\selectlanguage{english}
\title{Infinite-range correlations in 1D systems with continuous symmetry}


\author{Tobias Rindlisbacher}
\email{tobias.rindlisbacher@helsinki.fi}
\affiliation{University of Helsinki, Department of Physics,
P.O. Box 64, FI-00014 University of Helsinki, Finland}

\begin{abstract}
$\On{N}$-symmetric lattice scalar fields are considered, coupled to a chemical potential and source terms. At the example of $N=2$, it is shown that such systems can even in (0+1) dimensions produce infinite-range correlations and a non-zero vacuum expectation value whenever the chemical potential assumes certain discrete values. Different mechanisms for how the latter phenomena are produced are discussed, depending on whether source terms are set to zero or non-zero values. In the conclusion, the relation of these findings to the Mermin-Wagner theorem is addressed.
\end{abstract}

\keywords{\small $\On{N}$ scalar field; chemical potential; finite density; long-range order; Mermin-Wagner theorem.}

\maketitle


\clearpage
\section{Introduction}\label{sec:introduction}
Mermin and Wagner showed in 1966~\cite{Mermin}, using Bogoliubov's inequality~\cite{Bogoliubov}, that in one and two dimensions the ferromagnetic (antiferromagnetic) isotropic Heisenberg model with finite-range interaction cannot undergo spontaneous magnetization (sub-lattice magnetization) at any non-zero temperature. Isotropic here refers to the internal space in which the Heisenberg spins $\bar{s}_{x}=\ssof{s_{x}^{1},s_{x}^{2},s_{x}^{3}}$ take value, and means that if we write the interaction between two spins $\bar{s}_{x}$, $\bar{s}_{y}$, located on sites $x$ and $y$, as $\sum_{i=1}^{3}\,\alpha_{i}\of{x-y}\,s_{x}^{i}\,s_{y}^{i}$, then $\alpha_{1}\of{r}=\alpha_{2}\of{r}=\alpha_{3}\of{r}$ for all $r$. If, on the other hand, the couplings satisfy $\alpha_{1}\of{r}=\alpha_{2}\of{r}\neq \alpha_{3}\of{r}$, so that the global $\On{3}$-symmetry of the isotropic case is reduced to $\On{2}$, the authors mention that the same line of reasoning would then only rule out spontaneous magnetization (sub-lattice magnetization) in the $s^{1}$-$s^{2}$-plane (where it would break the reduced $\On{2}$-symmetry). The authors also stressed, that the impossibility of spontaneous magnetisation does not exclude the existence of other types of phase transitions in these models. For the theorem to apply, it is important that long-range interactions are sufficiently suppresssed~\cite{Thouless,Imry}.\\
In 1967 Wegner~\cite{Wegner} confirmed this for the ferromagnetic Heisenberg model with reduced $\On{2}$-symmetry ($\sim$XY-model), by showing, based on a low-temperature expansion, that the spin-spin correlation function in one and two dimensions always converges to zero at sufficiently large distances. The Mermin-Wagner theorem has in the following years been generalized to many other (non-relativistic) classical and quantum systems, cf.~\cite{Hohenberg,Hamilton,Moore,Mermin2,Berezinsky1,Berezinsky2,Kishore}, to mention only a few. In 1973 Coleman~\cite{Coleman} then proved the analogue of the Mermin-Wager theorem~\cite{Mermin} and the work of Wegner~\cite{Wegner} and Berezinsky~\cite{Berezinsky1,Berezinsky2} for a relativistic scalar field theory; showing that in two (i.e. $\ssof{1+1}$) dimensions, the vacuum expectation value of a scalar field with a continuous global symmetry is always zero, so that no Goldstone bosons can occur.\\

In this article, we consider non-linear $\On{N}$ lattice models in the presence of a chemical potential that couples to the conserved charge of a $\Un{1}$ sub-symmetry of $\On{N}$. On an Euclidean lattice, one can in these models for a discrete but infinite set of values of the chemical potential even in the one-dimensional case (i.e. in (0+1) dimensions) observe infinite range correlations and the formation of a non-zero vacuum expectation value. From a field theoretic point of view, this (0+1)-dimensional case is, of course, not particularly interesting, and it does also not conflict with Coleman's "no Goldstone bosons" theorem, as the latter addresses only $\ssof{1+1}$ dimensional systems. However, the Euclidean lattice formulations of our $\On{N}$ spin models can also be interpreted in a solid state physics context, with the the Euclidean action $S$ playing the role of $\beta\,H$, i.e. of the product of the inverse temperature $\beta=1/\ssof{k_{B}\,T}$ and the Hamiltonian $H$. From this point of view, the formation of long-range order in a one-dimensional system seems to conflict with the Mermin-Wagner theorem. As the one-dimensional lattice is in this case spatial, the chemical potential parameter $\mu$ couples to a spatial current and could be interpreted as a \emph{negative resistance}.\\

The paper is organized as follows: in the following section, Sec.~\ref{sec:model}, the lattice model as well as its formulation in terms of dual flux-variables will be introduced in detail. In Sec.~\ref{sec:results} we then show analytically that on a one-dimensional lattice, the model can for a discrete but infinite set of values of the chemical potential develop infinite range correlations and a non-zero vacuum expectation value. Sec.~\ref{sec:discussion} summarizes the findings and discusses their relation to the Mermin-Wagner theorem~\cite{Mermin}.\\       

\section{The model}\label{sec:model}
The model of interest to us in the present work is the non-linear $\On{N}$ spin model on a 1D (one-dimensional) Euclidean lattice, coupled to a chemical potential and source terms. In order to get an idea of how the lattice model is related to the corresponding continuum theory, we will first write down the action in $d=\ssof{d_{s}+1}$-dimensional continuous Minkowski space ($d_{s}$ being the number of spatial dimensions). We then go through the steps of performing the Wick-rotation to Euclidean time, and then putting the model on a $d=\ssof{d_{s}+1}$-dimensional Euclidean lattice. In this way, we can keep track of how the lattice parameters are related to their continuum counterparts and the lattice spacing $a$.\\
So, in $\ssof{d_{s}+1}$-dimensional Minkowski space, the action for our non-linear $\On{N}$-model reads:
\begin{widetext}
\[
S_{M}\ssfof{\phi}\,=\,\int\idd{x}{d}\,\sbcof{f_{\pi}^{2}\,\sof{\of{\partial_{\nu}+2\,\ii\,\mu_{c}\,\tau_{1 2}\,\delta^{0}_{\nu}}\phi_{c}\of{x}}^{\top}\!g^{\nu\rho}\,\sof{\of{\partial_{\rho}+2\,\ii\,\mu_{c}\,\tau_{1 2}\,\delta^{0}_{\rho}}\phi_{c}\of{x}}\,-\,j\of{x}\phi_{c}\of{x}}\ ,\label{eq:oncontinuumaction}
\]
\end{widetext}
where $g^{\nu\rho}=\diag\ssof{1,-1,\ldots,-1}$ is the Minkowski space metric, $\phi_{c}\in\Sph{N-1}\subset\mathbb{R}^{N}$, and $f_{\pi}$ is a (coupling) constant of mass dimension $\ssof{d-2}/2$, necessary to render the action (in natural units) dimension less while $\phi\in\Sph{N-1}$ is itself dimensionless. The field $j=\ssof{j^{1},\ldots,j^{N}}\in\mathbb{R}^{N}$ is a $N$-vector of source terms. The quantity $\tau_{1 2}$ that comes with the chemical potential $\mu_{c}$ is an $\On{N}$-generator $\ssof{\tau_{i j}}\indices{^a_b}=-\ii\ssof{\delta_{i}^{a}\delta_{j,b}-\delta_{j}^{a}\delta_{i,b}}$, so that $\mu_{c}$ couples to the conserved charge, corresponding to the $\Un{1}$-symmetry of rotations in the $\phi_{c}^{1}$-$\phi_{c}^{2}$-plane\footnote{The $2$ in front of the chemical potential in \eqref{eq:oncontinuumaction} is convention.}. The subscript c (for continuum) in $\phi_{c}$ and $\mu_{c}$ is just there to avoid confusing these continuum quantities with their lattice analogues that will be introduced shortly. If the chemical potential is non-zero, the global $\On{N}$-symmetry is explicitly broken to $\On{N-2}\!\times\!\Un{1}$. By setting $j\of{x}=J$, with $J\in\mathbb{R}^{N}$ to a position-independent, non-zero value, the symmetry can be reduced further. Setting the components $J^{1}$ or $J^{2}$ to non-zero values can be problematic in combination with a non-zero chemical potential, as these source components explicitly break the $\Un{1}$ symmetry in the $\phi_{c}^{1}$-$\phi_{c}^{2}$-plane and thereby render the charge to which the chemical potential couples non-conserved. However, as long as $\tilde{J}=\sqrt{\ssof{J^{1}}^2+\ssof{J^{2}}^2}$ is small, the $\Un{1}$-charge can still be considered approximately conserved. Non-zero values for the components $J^{i}$ with $i\geq 3$ are on the other hand never a problem and will just break the $\On{N-2}$ part of the global symmetry further down to $\On{N-3}$.\\
We now perform a Wick rotation, so that $\ii\,S_{M}\ssfof{\phi}\to -S_{E}\ssfof{\phi}$. With our sign convention for the Minkowski metric, this is achieved by setting $x^{0}=-\ii\,x^{d}$ (implying $\dd x^{0} = -\ii\,\dd x^{d}$ and $\partial_{0}=\ii\,\partial_{d}$), and the Euclidean action is then obtained as:
\begin{widetext}
\[
S_{E}\ssfof{\phi}\,=\,\int\idd{x}{d}\,\sbcof{f_{\pi}^{2}\,\sof{\of{\partial_{\nu}+2\,\mu\,\tau_{1 2}\,\delta_{\nu,d}}\phi\of{x}}^{\top}\!g^{\nu\rho}\,\sof{\of{\partial_{\rho}+2\,\mu\,\tau_{1 2}\,\delta_{\rho,d}}\phi\of{x}}\,+\,j\of{x}\phi\of{x}}\ ,\label{eq:oneuclcontinuumaction}
\]
\end{widetext}
where $g^{\nu\rho}$ and $\phi\of{x}$ are now the Euclidean metric and field, respectively, with the latter being obtained from the Minkowski space field by substituting $\phi\ssof{-\ii\,x^{d},x^{1},\ldots,x^{d-1}}\to\phi\ssof{x^{1},\ldots,x^{d}}$.\\
Finally, we put the theory on a lattice with finite lattice spacing $a$, substituting $\int\idd{x}{d}\,\to\,a^{d}\,\sum\limits_{x}$ and
\begin{multline}
\of{\partial_{\nu}+2\,\mu_{c}\,\tau_{1 2}\,\delta^{d}_{\nu}}\phi_{c}\of{x} \to\\
\frac{1}{a}\of{\phi_{x+\hat{\nu}}-\e^{-2\,\mu\,\tau_{1 2}\,\delta_{\nu,d}}\,\phi_{x}}\ ,\label{eq:discretederiv}
\end{multline}
where on the right-hand side of \eqref{eq:discretederiv}, factors of $a$, resp. $1/a$ have been absorbed into $\mu$ and $x$ by redefining $a\,\mu_{c}\to\mu$ and $x/a\to x$, in order to render the quantities dimensionless. The coordinate $x$ on the right-hand side of \eqref{eq:discretederiv} takes then values in $\mathbb{Z}^{d}$, and the relation between $\phi$ and $\phi_{c}$ is given by $\phi_{x}=\phi_{c}\ssof{a\,x}$ and $\phi_{x+\hat{\nu}}=\phi_{c}\ssof{a\ssof{x+e_{\nu}}}$ with $e_{\nu}$ being the unit vector in $\nu$-direction. After setting also $\beta=f_{\pi}^{2}\,a^{\ssof{d-2}}$ and $s=J\,a^{d}$, the Euclidean lattice action then takes the standard form:
\begin{widetext}
\[
S\fof{\phi}\,=\,-\sum\limits_{x}\bcof{\frac{\beta}{2}\,\sum\limits_{\nu=1}^{d}\of{\phi_{x}\,\e^{2\,\mu\,\tau_{1 2}\,\delta_{\nu,d}}\,\phi_{x+\hat{\nu}}+\phi_{x}\,\e^{-2\,\mu\,\tau_{1 2}\,\delta_{\nu,d}}\,\phi_{x-\hat{\nu}}}+\of{s\cdot\phi_{x}}}\ .\label{eq:onaction}
\]
\end{widetext}
Note, that for $d=\ssof{d_{s}+1}=1$ (i.e. in the case of spatial dimensionality $d_{s}=0$), the parameter $f_{\pi}$ has mass dimension -1/2 and its lattice counterpart is given by $\beta=f_{\pi}^{2}/a$. If $f_{\pi}$ is non-zero, we therefore have, that $\ssof{a\to 0}$ implies that $\ssof{\beta\to\infty}$, i.e. the parameter $\beta$ has to diverge in the continuum limit.

\subsection{Dual formulation}
The action \eqref{eq:onaction} is in general complex for non-zero values of the chemical potential $\mu$, which gives rise to a so-called \emph{sign-problem} when trying to evaluate the partition function,
\[
Z\,=\,\int\DD{\phi}\,\e^{-S\fof{\phi}}\ ,\label{eq:onpartfg}
\]
numerically by means of Monte Carlo simulations: a complex action $S$ implies that the Boltzmann weight $\e^{-S}$ is complex as well and therefore lacks the probabilistic interpretation, required to do importance sampling. This problem can be overcome by changing representation and expressing \eqref{eq:onpartfg} in terms of new, discrete variables. Such a change of representation is carried out in detail in appendix~\ref{ssec:dualizationofpartf}, following the steps described in \cite{Endres,Bruckmann,Rindlisbacher1,Rindlisbacher3}. After dualization, the partition function \eqref{eq:onpartfg} reads:
\begin{widetext}
\begin{multline}
Z\,=\,\sum\limits_{\cof{k,l,\chi,p,q,n}}\prod\limits_{x}\bcof{\bof{\prod\limits_{\nu=1}^{d}\frac{\beta^{\abs{k_{x,\nu}}+2\,l_{x,\nu}+\sum_{i=3}^{N}\,\chi^{i}_{x,\nu}}}{\of{\abs{k_{x,\nu}}+l_{x,\nu}}!\,l_{x,\nu}!\,\prod_{i=3}^{N}\chi^{i}_{x,\nu}!}}\\
\cdot\frac{\ssof{s^{+}}^{\frac{1}{2}\of{\abs{p_{x}}+p_{x}}+q_{x}}\ssof{s^{-}}^{\frac{1}{2}\of{\abs{p_{x}}-p_{x}}+q_{x}}\,\e^{2\,\mu\,k_{x,d}}}{\of{\abs{p_{x}}+q_{x}}!\,q_{x}!}\bof{\prod\limits_{i=3}^{N}\frac{\ssof{s^{i}}^{n^{i}_{x}}}{n^{i}_{x}!}}\,\delta\sof{p_{x}-\sum\limits_{\nu}\sof{k_{x,\nu}-k_{x-\hat{\nu},\nu}}}\\
\cdot W\sof{A_{x}+\abs{p_{x}}+2\,q_{x},\,B^{3}_{x}+n^{3}_{x},\,\ldots,\,B^{N}_{x}+n^{N}_{x}}}\ ,\label{eq:onfluxreppartfg}
\end{multline}
with
\[
A_{x}=\sum\limits_{\nu}\of{\sabs{k_{x,\nu}}+\sabs{k_{x-\hat{\nu},\nu}}+2\sof{l_{x,\nu}+l_{x-\hat{\nu},\nu}}}\quad,\quad B^{i}_{x}=\sum\limits_{\nu}\of{\chi^{i}_{x,\nu}+\chi^{i}_{x-\hat{\nu},\nu}}
\]
and
\[
W\sof{A,B^{3},\ldots,B^{N}}=\frac{\Gamma\sof{\frac{2+A}{2}}\,\prod\limits_{i=3}^{N}\frac{1+\of{-1}^{B^{i}}}{2}\Gamma\sof{\frac{1+\vphantom{B}\smash{B^{i}}}{2}}}{2^{\sfrac{A}{2}}\,\Gamma\sof{\frac{N+A+\sum_{i=3}^{N}B^{i}}{2}}}\ ,\label{eq:onweightfuncg}
\]
\end{widetext} 
and where, for all sites $x$ and all links $\of{x,\nu}$:
\begin{subequations}
\begin{align}
k_{x,\nu},p_{x}&\in\mathbb{Z}\ ,\\ 
l_{x,\nu},q_{x},\chi^{3}_{x,\nu},n^{3}_{x},\ldots,\chi^{N}_{x,\nu},n^{N}_{x}&\in\mathbb{N}_{0}\ .
\end{align}
\end{subequations}
The delta function at the end of the second line of \eqref{eq:onfluxreppartfg} is just a Kronecker delta, $\delta\of{x}=\delta_{0,x}$, and the new parameters $s^{\pm}=\frac{1}{\sqrt{2}}\ssof{s^{1}\mp\ii\,s^{2}}$ are the sources for the $\Un{1}$-charged $\phi^{\pm}=\frac{1}{\sqrt{2}}\of{\phi^{1}\pm\ii\,\phi^{2}}$.\\
While in \eqref{eq:onpartfg} the partition function was given in terms of an integral over a continuum of configurations $\cof{\phi_{x}}_{x}$ with complex weights, in the flux-representation \eqref{eq:onfluxreppartfg} it is given in terms of an infinite sum over configurations of discrete variables $\sscof{k,l,\chi^{i},p,q,n^{i}}$, which all have real and non-negative weights\footnote{For a configuration to have a real and non-negative weight, the weight factor $W_{s}=\prod_{x}\ssof{s^{+}}^{\frac{1}{2}\of{\abs{p_{x}}+p_{x}}+q_{x}}\ssof{s^{-}}^{\frac{1}{2}\of{\abs{p_{x}}-p_{x}}+q_{x}}$ has to be real and non-negative. That this is the case can be seen, by writing $s^{\pm}=\frac{\tilde{s}\,\e^{\mp\ii\phi_{s}}}{\sqrt{2}}$, so that $W_{s}=\tilde{s}^{\sum_{x}\ssof{\abs{p_{x}}+2\,q_{x}}}\,\e^{\ii\phi_{s}\sum_{x}\,p_{x}}$, and noting that due to the Kronecker delta on the second line of \eqref{eq:onfluxreppartfg}, we have $\sum_{x}\,p_{x}=\sum_{x,\nu}\of{k_{x,\nu}-k_{x-\hat{\nu},\nu}}=0$}.\\
In the present work we will primarily be interested in the $\of{0+1}$-dimensional case, which could also in the standard formulation be solved using transfer-matrix methods, regardless of whether the action is real or complex. The reformulation of the partition function in terms of flux-variables is nevertheless useful as it simplifies the necessary computations considerably. Even more important is, however, that as discussed in the next section, the flux-variables have a direct physical meaning, which can be useful in identifying processes underlying different thermodynamic properties.

\subsection{Physical meaning of dual variables}\label{ssec:physmeaningoffluxvar}
The variables $k$, $l$ and $\chi^{i}$ live on the links of the lattice and are called \emph{flux-variables}. They can in general be interpreted as counting the number of particle world lines passing along the links. So, for example $\chi^{i}_{x,\nu}$, with $i\in\cof{3,\ldots,N}$, counts the number of $\phi^{i}$-world lines traversing the link that connects the sites $x$ and $x+\hat{\nu}$. Because the $\phi^{i}$-particles (with $i\in\cof{3,\ldots,N}$) are neutral, they are their own anti-particles, so that their world lines don't need an orientation. For the $\phi^{\pm}$ particles, which carry a $\Un{1}$-charge, the situation is different: it matters whether a $\phi^{+}$ moves from $x$ to $x+\hat{\nu}$ or from $x+\hat{\nu}$ to $x$, as the two cases lead to different charge displacements. On the other hand, it doesn't make a difference whether a $\phi^{+}$ moves from $x$ to $x+\hat{\nu}$, or a $\phi^{-}$ moves form $x+\hat{\nu}$ to $x$, as the charge-displacement is in both cases the same ($\mathcal{CT}$ symmetry is preserved). The $k$ and $l$ variables are therefore picked in such a way, that $k_{x,\nu}$ counts the net-positive charge that moves from site $x$ to site $x+\hat{\nu}$, whereas $l_{x,\nu}$ counts the number of neutral pairs of $\phi^{\pm}$ world lines that pass along the link between the two sites.\\
The remaining variables $p$, $q$ and $n^{i}$, with $i\in\cof{3,\ldots,N}$, live on the lattice sites and are called \emph{monomer numbers}. In analogy to the flux-variables, the monomer numbers are organized so that $p_{x}$ counts the net-positive charge on site $x$ and $q_{x}$ the number of neutral pairs of $\phi^{\pm}$-monomers, while $n^{i}_{x}$ counts the number of $\phi^{i}$-monomers on site $x$.\\
Due to the delta-function constraints in \eqref{eq:onfluxreppartfg}, the $k$- and $p$-variables are on each site $x$ subject to a conservation law:
\[
\sum_{\nu}\of{k_{x,\nu}-k_{x-\hat{\nu},\nu}}\,\mbeq\,p_{x}\ ,\label{eq:lattgausslawg}
\]
which is a lattice manifestation of Gauss' law $\partial_{\nu}j^{\nu}\of{x}=\rho_{s}\of{x}$, with $k_{x,\nu}$ being related to the current density, $j^{\nu}\of{x}=\ii\,\of{\phi^{+}\of{x}\ssof{\partial^{\nu}\phi^{-}\of{x}}-\phi^{-}\of{x}\ssof{\partial^{\nu}\phi^{+}\of{x}}}$, and $p_{x}$ (which can be non-zero only if $\abs{s^{\pm}}>0$) being related to the $\Un{1}$-breaking field $\rho_{s}\of{x}=\ii\,\ssof{s^{+}\,\phi^{+}-s^{-}\,\phi^{-}}$. While showing, that for $\ssabs{s^{\pm}}>0$, charge conservation is locally violated, equation \eqref{eq:lattgausslawg} also implies that for periodic lattices, we have:
\[
\sum_{x}\,p_{x}\,=\,\sum_{x,\nu}\of{k_{x,\nu}-k_{x-\hat{\nu},\nu}}\,=\,0\ ,
\] 
meaning that the overall charge carried by $\phi^{\pm}$-monomers must be zero ($\sim$ integral-form of Gauss' law in a space without boundary). Note, however, that in this quantum field theory context, the number of $\phi^{\pm}$-monomers has nothing to do with the electric charge density in the system! The latter, which in the continuum would be given by $j^{0}$ (or $j^{d}$ in Euclidean space-time), is represented by the flux variables $k_{x,d}$ that live on the time-like links.\\
The $l$- and $q$-variables are completely unconstrained as they represent the numbers of neutral pairs of $\phi^{\pm}$-world line segments or monomers, which can at any point be created and annihilate. This is not the case for the $\chi^{i}$- and $n^{i}$-variables, which are on each site $x$ subject to the \emph{evenness-constraint},
\[
\sof{n^{i}_{x}\,+\,\sum_{\nu}\ssof{\chi^{i}_{x,\nu}-\chi^{i}_{x-\hat{\nu},\nu}}}\bmod 2\,\mbeq\,0\ ,
\]
encoded in the site weight \eqref{eq:onweightfuncg}. This constraint reflects the fact that also neutral particles can be created and annihilated only in pairs, so that a world line, that enters a site, has either to continue and leave the site, or to annihilate with a monomer.\\

\section{Results}\label{sec:results}
For simplicity we discuss in the following the case of $N=2$, which corresponds to an XY-model that is coupled to source terms and a chemical potential for the $\Un{1}$ charge. The qualitative findings generalize to $N>2$, as well as to the case of linear spin models with arbitrary on-site potentials.\\
With $N=2$, the site weight \eqref{eq:onweightfuncg} simplifies to $W\of{A}=2^{-A/2}$. The summations over the individual $l$- and $q$-variables in the partition function \eqref{eq:onfluxreppartfg} then decouple, and the partition function simplifies to:
\begin{widetext}
\[
Z\,=\,\sum\limits_{\cof{k,p}}\prod\limits_{x}\bcof{\bof{\prod\limits_{\nu=1}^{d}\,I_{k_{x,\nu}}\of{\beta}}\,I_{p_{x}}\of{\tilde{s}}\,\e^{\ii\,\phi_{s}\,p_{x}}\,\e^{2\,\mu\,k_{x,d}}\,\delta\sof{p_{x}-\sum\limits_{\nu}\sof{k_{x,\nu}-k_{x-\hat{\nu},\nu}}}}\ ,\label{eq:o2fluxreppartfg}
\]
\end{widetext}
with $I_{n}\of{\alpha}$ being the modified Bessel function of the first kind\footnote{The modified Bessel function of the first kind can be written as $I_{n}\of{\alpha}=\sum_{m=0}^{\infty}\frac{\of{\frac{\alpha}{2}}^{n+2\,m}}{\of{n+m}!\,m!}$, which is precisely the form of the sums over $l$ and $q$ in \eqref{eq:onfluxreppartfg} when $N=2$, with the replacements ($m\to\abs{k}$, $\alpha\to\beta$) and ($m\to\abs{p}$, $\alpha\to\tilde{s}$). Note also that the factorials in the denominator of the expression for the Bessel function are understood in terms of gamma-functions, i.e. $n!=\Gamma\of{1+n}$, so that $I_{n}\of{\alpha}=I_{\abs{n}}\of{\alpha}$ and we therefore don't have to write the modulus in the subscript of $I_{n}\of{\alpha}$.} and $\tilde{s}$ and $\phi_{s}$ are magnitude and phase of the sources, so that $s^{\pm}=\frac{\tilde{s}\,\e^{\mp\,\ii\,\phi_{s}}}{\sqrt{2}}$.\\

In (0+1) dimensions, i.e. when the lattice extends only in the Euclidean time direction, the partition function \eqref{eq:o2fluxreppartfg} simplifies further. Setting for the moment $\tilde{s}=0$ (which implies $p_{x}=0$), it reduces to a single sum:
\[
Z\,=\,\sum\limits_{k=-\infty}^{\infty}\of{I_{k}\of{\beta}\,\e^{2\,\mu\,k}}^{N_{t}}\ ,\label{eq:o2fluxreppartf1d}
\]
with $N_{t}$ being the temporal extent (number of sites) of the periodic lattice. The summation variable $k$ represents the collective value of all the $k_{x,d}$-variables in the system and each value of $k$ therefore represents a distinct homogeneous (in terms of flux variables) configuration. As discussed in Sec.~\ref{ssec:physmeaningoffluxvar}, the values of the $k_{x,d}$-variables can be identified with quanta of the local charge density $n=j^{d}$ in a dual configuration and the value of $k$ in \eqref{eq:o2fluxreppartf1d} is therefore not just a label, but represents the charge density carried by the corresponding homogeneous configuration. This can be verified by taking the derivative of the logarithm of the partition function with respect to $\mu$: in arbitrary dimensions, this yields the charge density:
\begin{multline}
\avof{n}\,=\,\frac{1}{2\,V}\partd{\log\of{Z}}{\mu}\\
=\,\frac{1}{V}\sum_{x}\avof{k_{x,d}}_{dual}\,\overset{\of{d=\ssof{0+1}}}{=}\,\avof{k}_{dual}\ ,\label{eq:o2fluxrepchargedens1d}
\end{multline}
and we see that in (0+1) dimensions, using the partition function \eqref{eq:o2fluxreppartf1d}, the result is given by the expectation value of $k$.\\
The partition function \eqref{eq:o2fluxreppartf1d} is therefore a sum over distinct homogeneous (in terms of dual variables) configurations, which are labelled by the integer-valued charge (density) $k$ that they carry.

\subsection{Transitions between distinct ground states}\label{ssec:phasetransitions}
\begin{figure}[ht]
\centering
\begin{minipage}[t]{0.49\textwidth}
\centering
\includegraphics[width=0.88\linewidth]{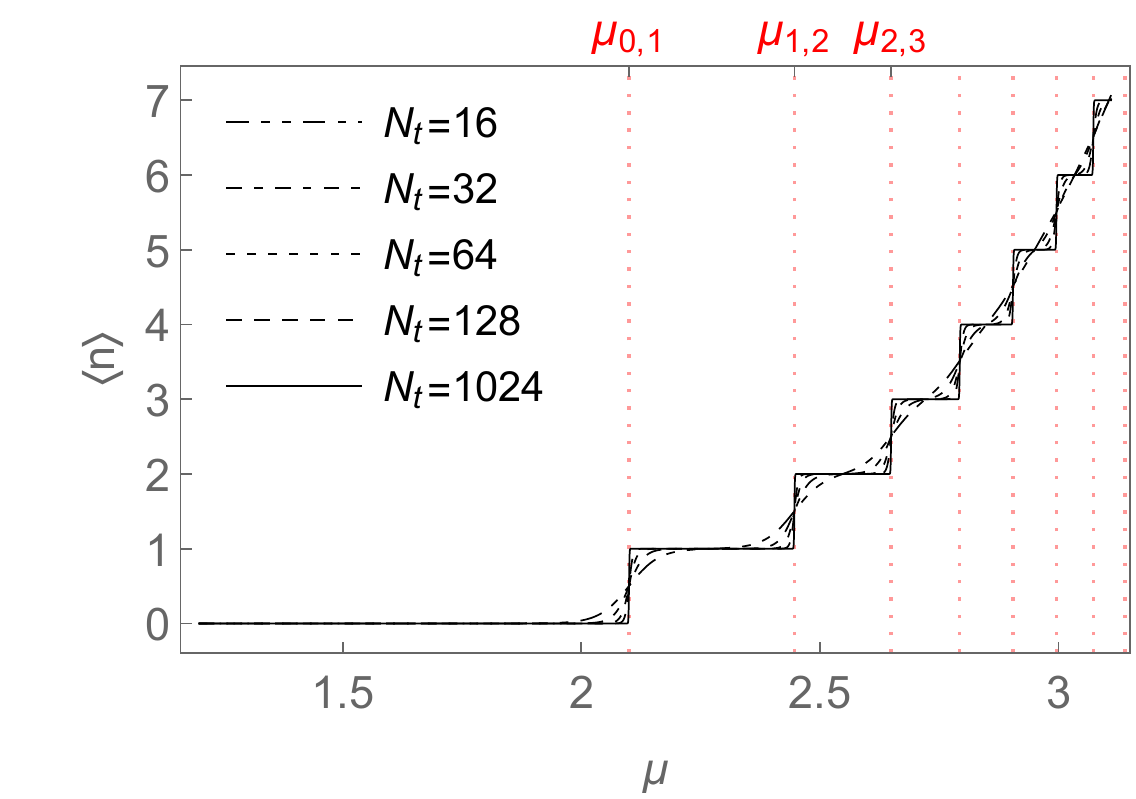}
\end{minipage}\\[10pt]
\begin{minipage}[t]{0.49\textwidth}
\centering
\includegraphics[width=0.88\linewidth]{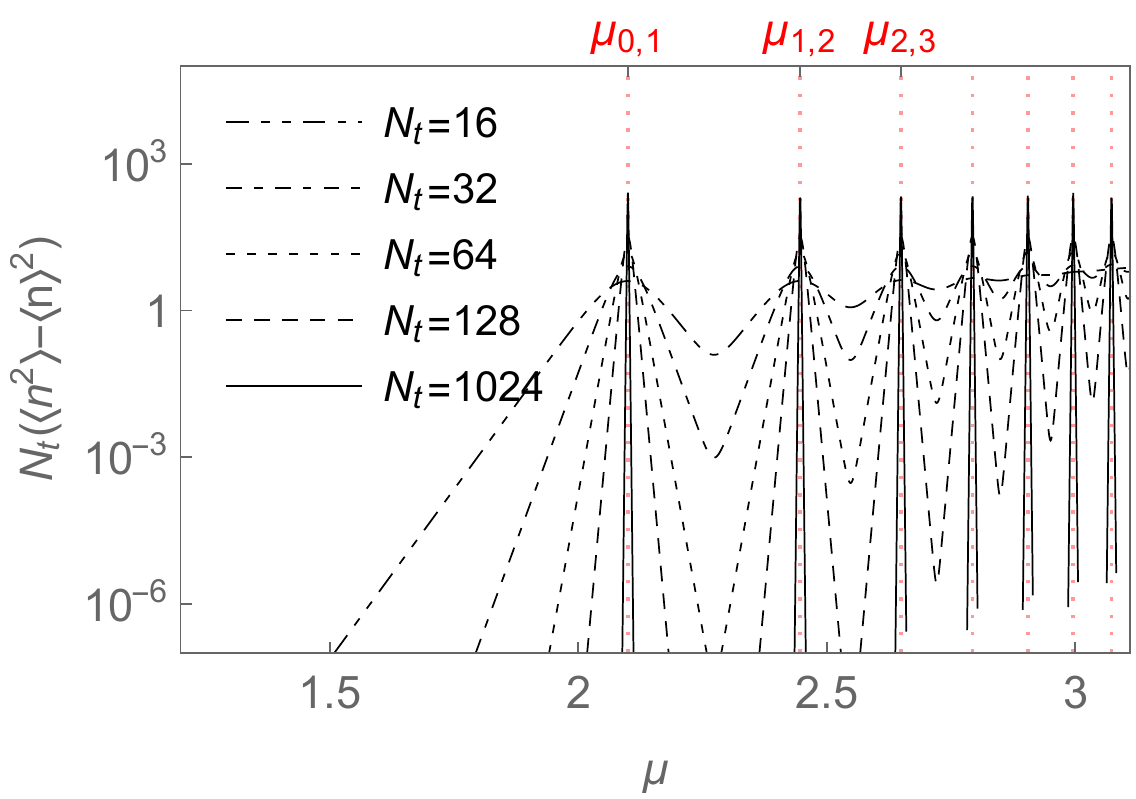}
\end{minipage}
\caption{The figure shows the charge density (top) and corresponding susceptibility (bottom) as functions of the chemical potential $\mu$ for the $\On{2}$-model in (0+1) dimensions for $\tilde{s}=0$ (c.f. eq. \eqref{eq:o2fluxreppartf1d}), $\beta=0.03$ and lattice sizes $N_{t}=$16, 32, 64, 128, 1024. At subsequent critical values of $\mu=\mu_{0,1}$, $\mu_{1,2}$, $\ldots$, the system undergoes transitions between vacua carrying different integer-valued charge densities. The transitions become apparently stronger with increasing system size, as one would expect from \eqref{eq:o2fluxrepchargedens1d}.}
\label{fig:densvsmud1s0}
\end{figure}

For each fixed value of $\beta$, the modified Bessel function of the first kind, $I_{n}\of{\beta}$, decays super-exponentially as function of increasing $\abs{n}$ (otherwise, the sum in \eqref{eq:o2fluxreppartf1d} would diverge), and it is therefore evident, that for $\mu=0$, the term in \eqref{eq:o2fluxreppartf1d} with $k=0$ dominates if $N_{t}$ is sufficiently large. With increasing $\mu$, however, the factor $\e^{2\,\mu\,k}$ will at some point be able to compensate for the decrease in $I_{k}\of{\beta}$ when $k$ changes from zero to one, and if $\mu$ exceeds this point, the term in \eqref{eq:o2fluxrepchargedens1d} with $k=1$ will become the dominant one. As all the configurations with different $k$ are homogeneous configurations (in terms of our flux variables), each of them qualifies in the $\ssof{N_{t}\to\infty}$-limit as vacuum state if its weight dominates the sum in \eqref{eq:o2fluxrepchargedens1d}. The transitions between states with different index $k$ can therefore be understood as transitions between different ground states which carry different integer-valued charge density. These transitions occur whenever the value of $\mu$ is such that two successive terms in \eqref{eq:o2fluxrepchargedens1d} become degenerate, i.e. when we have for some $k\in\mathbb{Z}$, that $\mu$ is such that:
\[
I_{k}\of{\beta}\,\e^{2\,\mu\,k}\,=\,I_{k+1}\of{\beta}\,\e^{2\,\mu\,\of{k+1}}\ .
\]
Upon solving for $\mu$, this yields the critical value  
\[
\mu_{k,k+1}\,=\,\frac{1}{2}\log\of{\frac{I_{k}\of{\beta}}{I_{k+1}\of{\beta}}}\ ,\label{eq:o2critmu1d}
\]
for the transition between the states carrying charge densities $k$ and $k+1$, respectively. That the system described by \eqref{eq:o2fluxreppartf1d} indeed undergoes such transitions between successive discrete states when the chemical potential is increased, is illustrated in in the upper panel of Fig.~\ref{fig:densvsmud1s0} for systems of different size $N_{t}$ and with $\beta=0.03$. The quantity in the lower panel is the corresponding charge-susceptibility,
\begin{multline}
\chi_{n}\,=\,V\,\sof{\ssavof{n^{2}}-\ssavof{n}^{2}}\,=\,\frac{1}{4\,V}\,\partdm{\log\of{Z}}{\mu}{2}\\
=\,V\,\sof{\ssavof{k_{x,d}^{2}}_{dual}-\ssavof{k_{x,d}}_{dual}^{2}}\\
\overset{\of{d=1}}{=}\,N_{t}\sof{\ssavof{k^{2}}_{dual}-\ssavof{k}^{2}_{dual}}\ ,\label{eq:o2fluxrepchargesusc1d}
\end{multline}
which, with increasing $N_{t}$, becomes more and more peaked at the transition points $\mu_{k,k+1}$, $k\in\mathbb{Z}$.\\
\begin{figure}[h]
\centering
\begin{minipage}[t]{0.49\textwidth}
\centering
\includegraphics[width=0.85\linewidth]{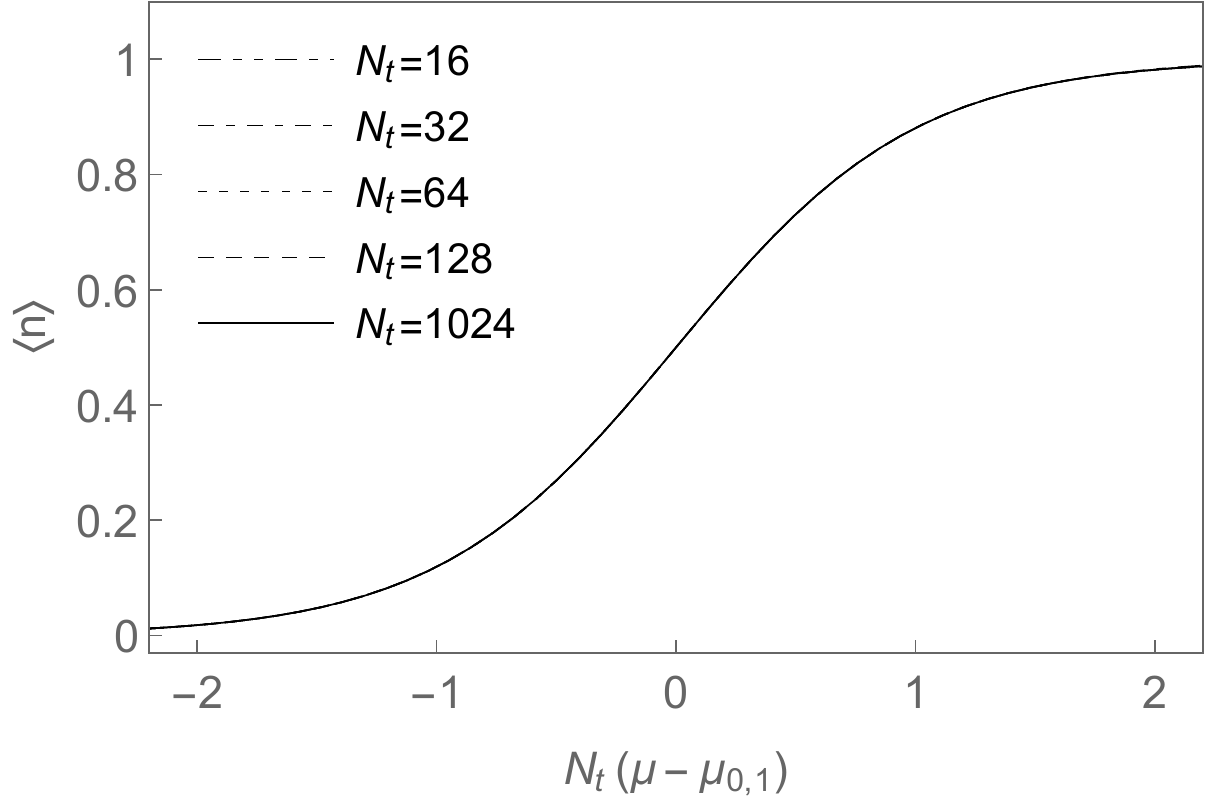}
\end{minipage}\\[5pt]
\begin{minipage}[t]{0.49\textwidth}
\centering
\includegraphics[width=0.85\linewidth]{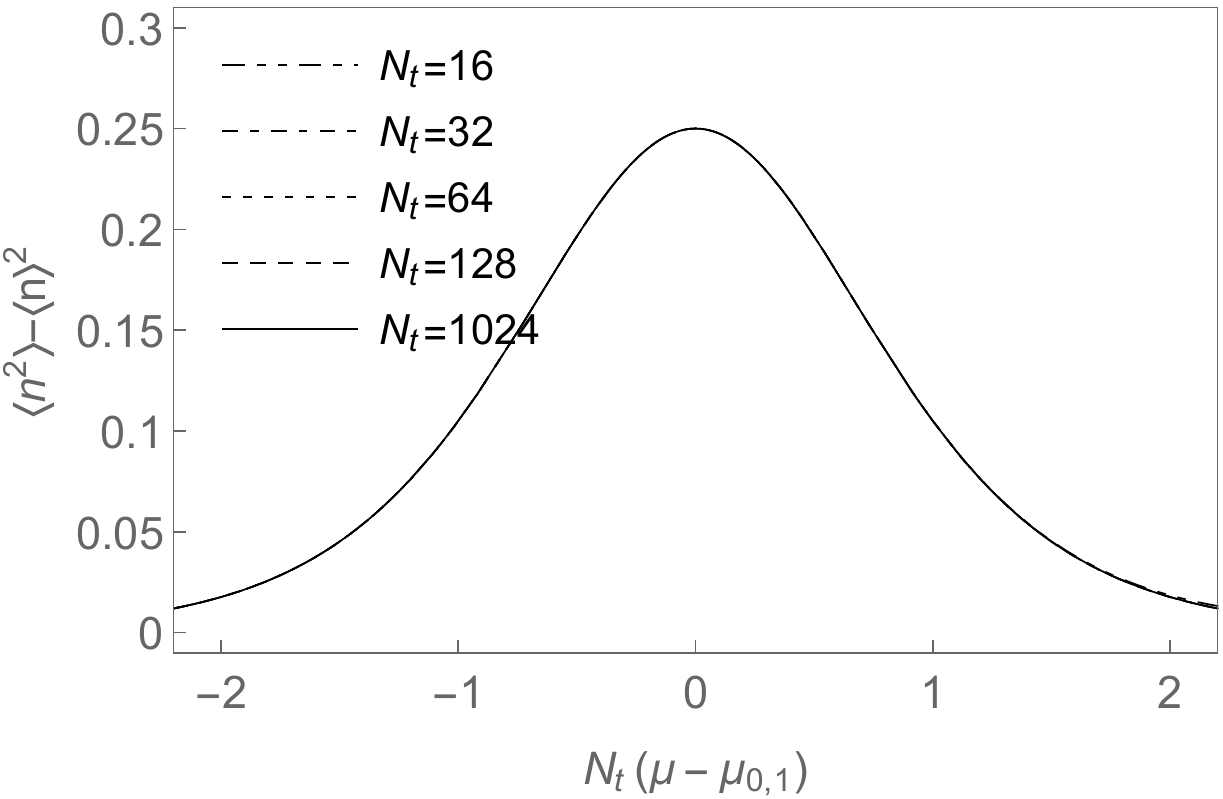}
\end{minipage}
\caption{The figure shows for the $\On{2}$-model in (0+1) dimensions with $\tilde{s}=0$ (c.f. eq. \eqref{eq:o2fluxreppartf1d}), $\beta=0.03$ and lattice sizes $N_{t}=16,32,64,128,1024$, scaling collapses for the charge density (top) and the charge susceptibility (bottom) around the first critical point $\mu=\mu_{0,1}$. The collapses are obtained according to \eqref{eq:o2fluxrepchargedense1drescaled} and \eqref{eq:o2fluxrepchargesusc1drescaled}.}
\label{fig:densvsmud1s0rescaleda}
\end{figure}

\begin{figure}[h]
\centering
\begin{minipage}[t]{0.49\textwidth}
\centering
\includegraphics[width=0.85\linewidth]{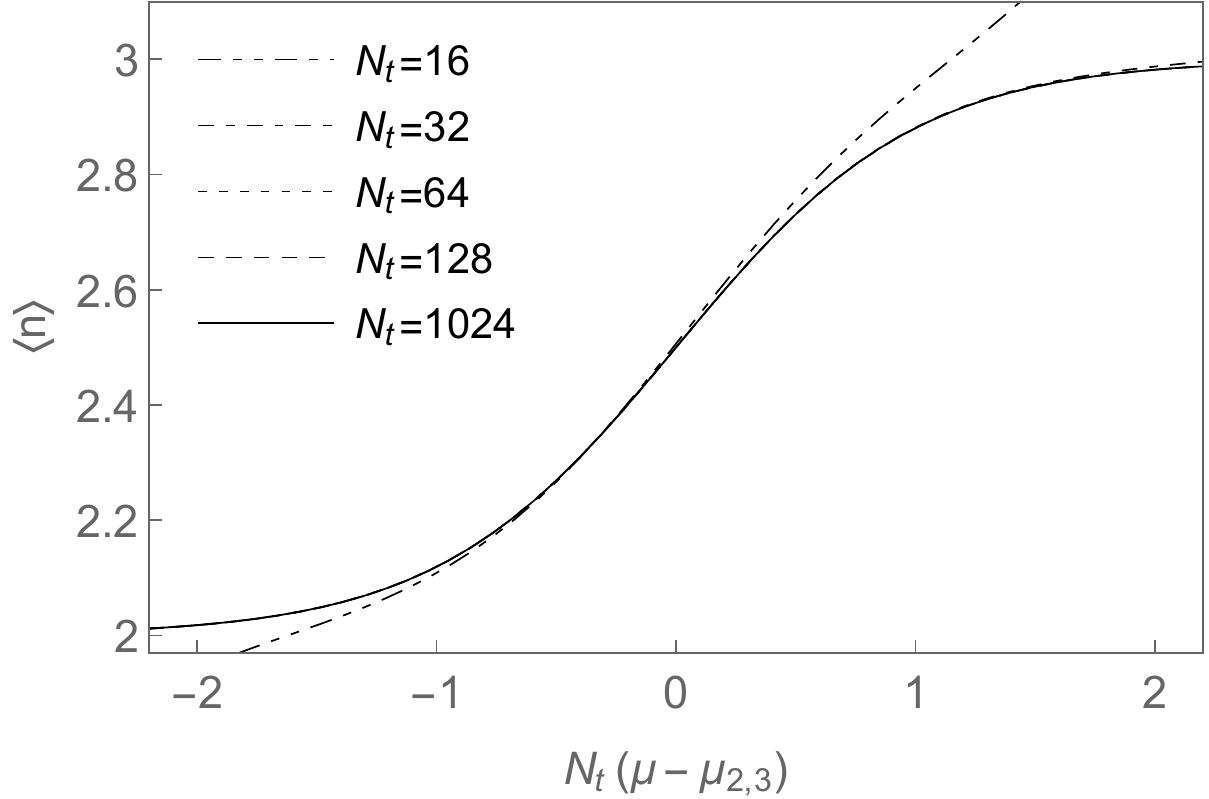}
\end{minipage}\\[5pt]
\begin{minipage}[t]{0.49\textwidth}
\centering
\includegraphics[width=0.85\textwidth]{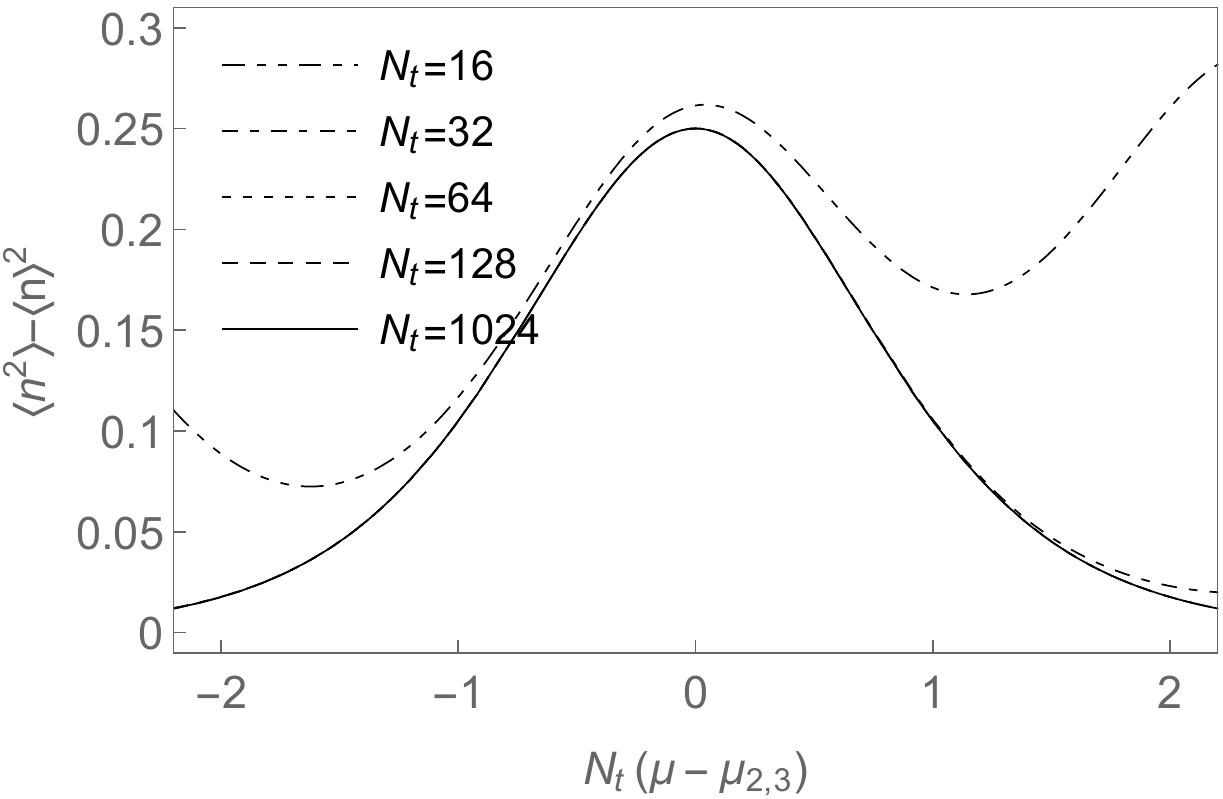}
\end{minipage}
\caption{Same as Fig.~\ref{fig:densvsmud1s0rescaleda}, but here the scaling collapses are performed at the critical point $\mu=\mu_{2,3}$ instead of $\mu=\mu_{0,1}$.}
\label{fig:densvsmud1s0rescaledb}
\end{figure}

In the neighbourhood of each critical point $\mu_{k,k+1}$, one can rescale $\ssavof{n}\of{\mu,N_{t}}$ and $\chi_{n}\of{\mu,N_{t}}$, so that the curves for different $N_{t}$ coincide. This is achieved by setting for each $k$:
\[
\avof{\tilde{n}}_{k}\of{\Delta\tilde{\mu}}=\avof{n}\of{\mu_{k,k+1}+N_{t}^{-1}\Delta\tilde{\mu},N_{t}}\label{eq:o2fluxrepchargedense1drescaled}
\]
and
\[
\tilde{\chi}_{n,k}\of{\Delta\tilde{\mu}}=
N_{t}^{-1}\chi_{n}\of{\mu_{k,k+1}+N_{t}^{-1}\Delta\tilde{\mu},N_{t}}\ ,\label{eq:o2fluxrepchargesusc1drescaled}
\]
where for each $k$, $\Delta\tilde{\mu}$ is, in terms of the original, unscaled $\mu$, given by $\Delta\tilde{\mu}=N_{t}\ssof{\mu-\mu_{k,k+1}}$. Examples for such \emph{scaling collapses} are shown in Figs.~\ref{fig:densvsmud1s0rescaleda}-\ref{fig:densvsmud1s0rescaledb} for $\beta=0.03$ and $\tilde{s}=0$, i.e. for the same systems that were used in Fig.~\ref{fig:densvsmud1s0}. 
As the rescaled charge-susceptibility \eqref{eq:o2fluxrepchargesusc1drescaled} is independent of $N_{t}$, the form of the rescaling implies that the original $\chi_{n}$ converges towards a sum of Dirac delta-functions when the thermodynamic limit $\of{N_{t}\to\infty}$ is taken at fixed $\beta$ (corresponding to taking the zero-temperature limit at finite lattice spacing). The charge density $\avof{n}$ becomes therefore discontinuous in this limit and the transition is of first order.\\

\begin{figure}[ht]
\centering
\begin{minipage}[t]{0.49\textwidth}
\centering
\includegraphics[width=0.85\linewidth]{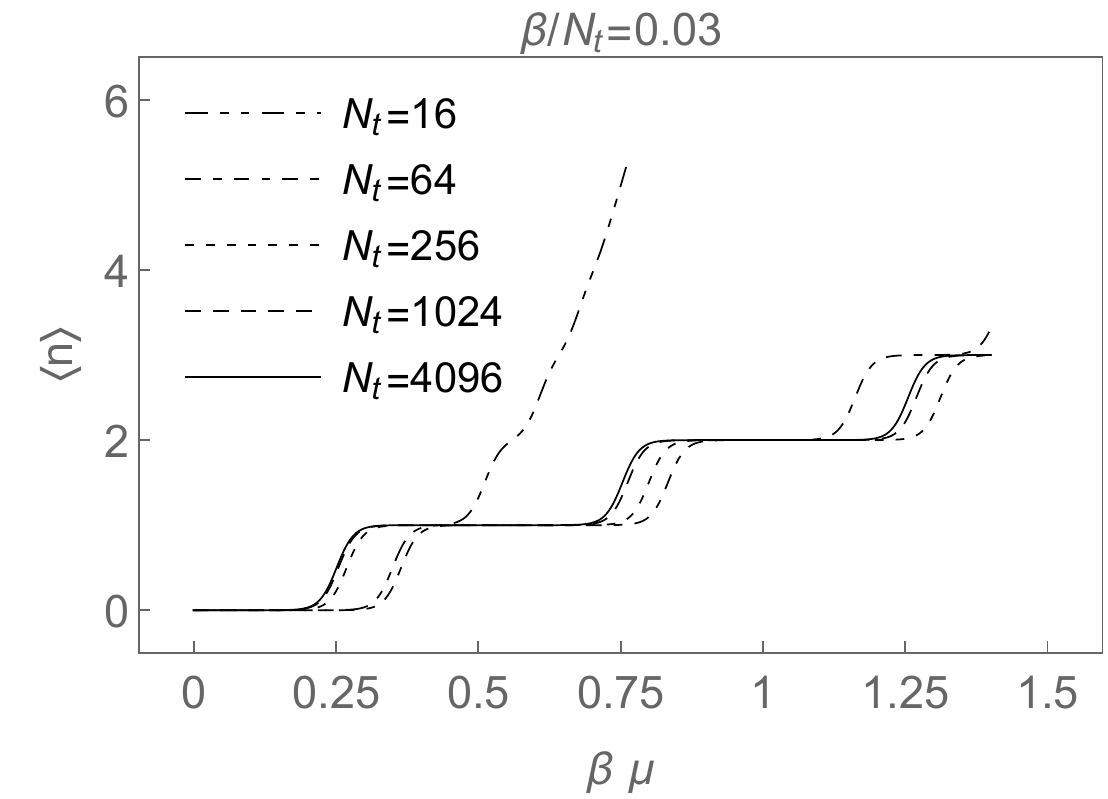}
\end{minipage}\\[5pt]
\begin{minipage}[t]{0.49\textwidth}
\centering
\includegraphics[width=0.85\linewidth]{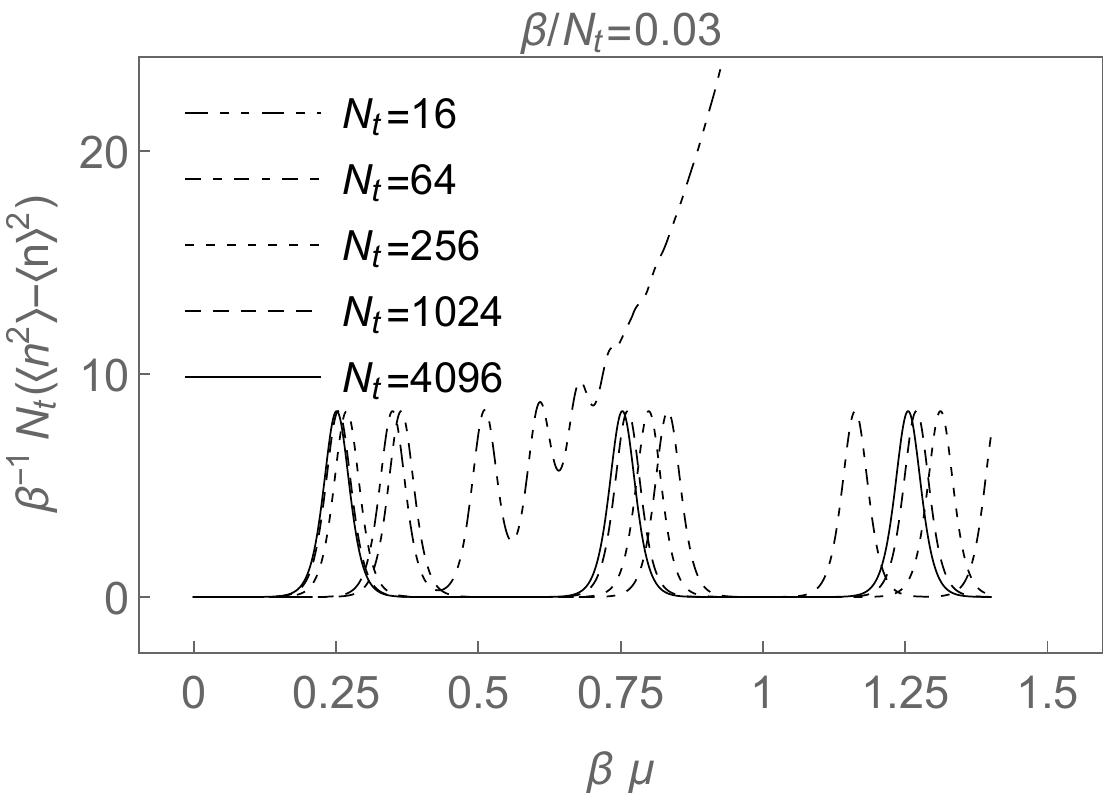}
\end{minipage}
\caption{The figure shows for the $\On{2}$ model in (0+1) dimensions the charge density (top) and charge susceptibility (bottom) as function of $\beta\,\mu$ at fixed $T\,f_{\pi}^{2}=\beta/N_{t}=0.03$ for various lattice sizes $N_{t}=16,64,256,1024,4096$. The continuum limit would be obtained by sending $N_{t}\to\infty$.}
\label{fig:densvshmud1s0a}
\end{figure}

\begin{figure}[ht]
\centering
\begin{minipage}[t]{0.49\textwidth}
\centering
\includegraphics[width=0.85\linewidth]{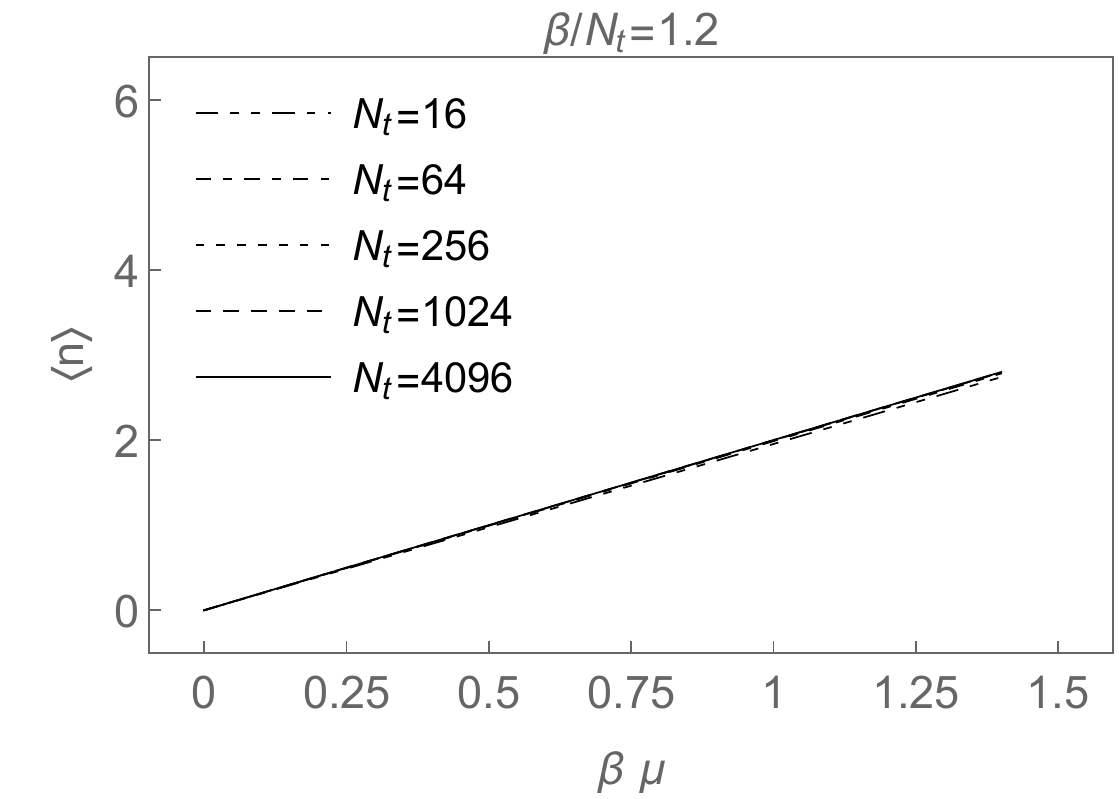}
\end{minipage}\\[5pt]
\begin{minipage}[t]{0.49\textwidth}
\centering
\includegraphics[width=0.85\linewidth]{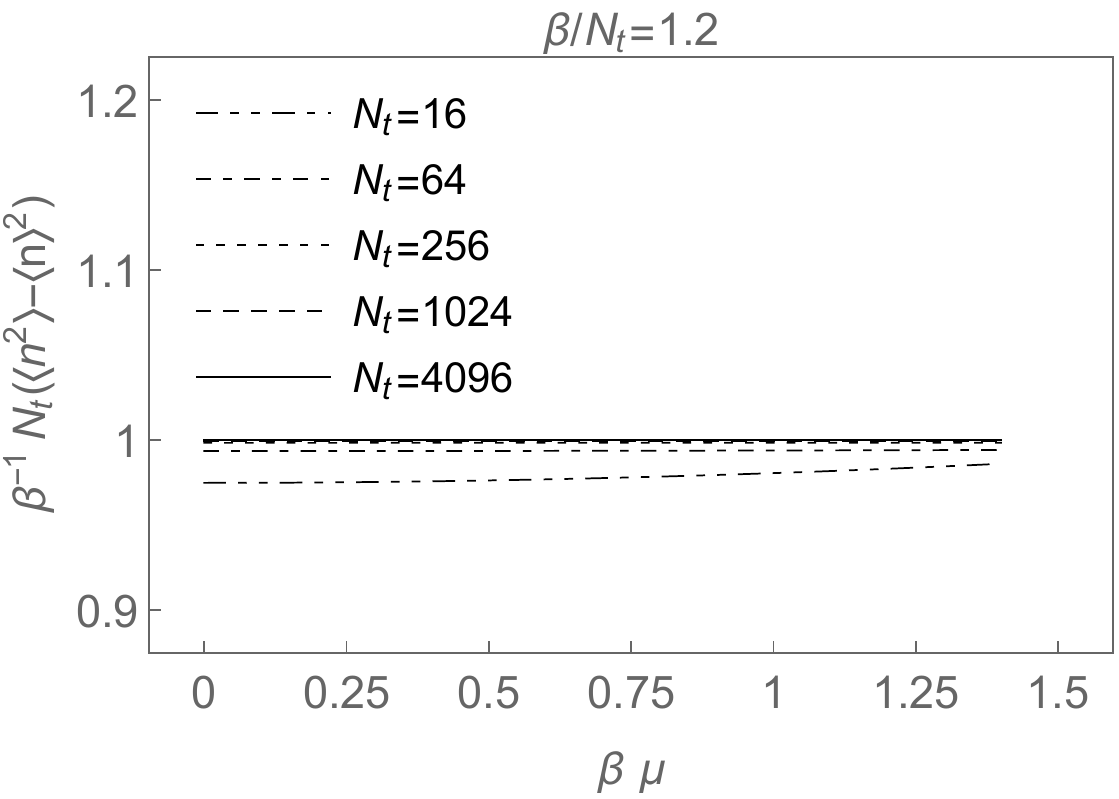}
\end{minipage}
\caption{Same as Fig.~\ref{fig:densvshmud1s0a} but at higher temperatures $T\,f_{\pi}^{2}=\beta/N_{t}=1.2$.}
\label{fig:densvshmud1s0b}  
\end{figure}

The situation changes if we take the thermodynamic limit not by sending the temperature to zero while keeping the lattice spacing fixed, but instead sending the lattice spacing to zero while keeping the temperature fixed. As $\beta$ is given by $\beta=f_{\pi}^{2}/a$, and $f_{\pi}$ has a physical meaning (i.e. it should assume a constant value if one enters the scaling window), we can do this by keeping $\kappa:=\beta/N_{t}=T\,f_{\pi}^{2}$ constant while sending $N_{t}$ to infinity. This is illustrated in Figs.~\ref{fig:densvshmud1s0a}-\ref{fig:densvshmud1s0b}, which show for different values of the temperature (i.e. different values of $\kappa=\beta/N_{t}$), how the charge density \eqref{eq:o2fluxrepchargedens1d} and the corresponding susceptibility \eqref{eq:o2fluxrepchargesusc1d} (multiplied by $\beta^{-1}$ to cancel the dependency on the lattice spacing) as functions of $\beta\,\mu$ behave when $N_{t}$ is changed: at low temperatures (cf. Fig.~\ref{fig:densvshmud1s0a}) one can again observe the formation of plateaus at integer values of the charge density and for $\ssof{N_{t}\to\infty}$ the curves converge towards a fixed shape that represents the continuum limit. In contrast to the situation we had in Fig.~\ref{fig:densvsmud1s0}, the charge density as function $\mu$ now remains smooth when $\kappa$ instead of $\beta$ is held fixed while $N_{t}$ is sent to infinity.\\
Although decreasing the value of $\kappa$ (lowering the temperature) makes the crossovers between the sectors of different integer-valued charge density become more abrupt, the crossovers are turned into first order transitions only if in the end also the zero-temperature limit is taken, i.e. when $\kappa$ is sent to zero. For increasing $\kappa$, on the other hand (cf. Fig.~\ref{fig:densvshmud1s0b}), the step-like behavior of the charge density as function of $\mu$ slowly disappears and turns into a linear behavior: $\ssavof{n}\approx 2\,\mu\,\beta$.\\

The $N_{t}$-dependency of charge density and charge susceptibility as function of $\mu$ at fixed $\kappa=\beta/N_{t}=T\,f_{\pi}^{2}$ can be understood from the behavior of $\beta\,\mu_{n,n+1}\ssof{\beta}$ as function of $\beta$. With the expression for $\mu_{n,n+1}\ssof{\beta}$ from eq. \eqref{eq:o2critmu1d}, one finds the asymptotic behavior of $\beta\,\mu_{n,n+1}\ssof{\beta}$ at small and large values of $\beta$ to be given by:
\[
\beta\,\mu_{n,n+1}\ssof{\beta}\,\approx\,\ucases{\frac{\beta}{2}\log\of{\frac{2\of{n+1}}{\beta}}\ \text{if}\ \beta \ll 1\\\frac{2\,n+1}{4}\ \text{if}\ \beta \gg 1}\ ,\label{eq:asymptoticbmucr}
\]
with a maximum separating the two regions. Clearly, for the curves in Figs.~\ref{fig:densvshmud1s0a}-\ref{fig:densvshmud1s0b} to converge, the $N_{t}$-values have to be sufficiently large, so that the corresponding values of $\beta\ssof{N_{t}}=\kappa\,N_{t}$ are all in the region $\beta\gg 1$, where $\beta\,\mu_{n,n+1}\ssof{\beta}$ is essentially independent of $\beta$. If the physical temperature is very low (i.e. if $\kappa$ is very small) then $\beta\ssof{N_{t}}$ can for small values of $N_{t}$ fall into the range $\beta\ll 1$, where $\beta\,\mu_{n,n+1}\of{\beta}$ grows only logarithmically instead of linearly with $n$. As function of $\beta\,\mu$, the successive transitions to higher and higher charge densities then happen for these small systems at much lower values of $\beta\,\mu$ than for the larger systems. This is the reason why in Fig.~\ref{fig:densvshmud1s0a} the charge density and susceptibility grow for $N_{t}=16$ much faster as function of $\beta\,\mu$ than for the larger values of $N_{t}$.\\

Finally, it can also be understood, why the step-like behavior of the charge density disappears when $\kappa=\beta/N_{t}$ is sufficiently large. To this end, we define the \emph{weight ratios},
\begin{multline}
W_{n_{1},n_{2}}\of{\mu,\beta,N_{t}}=\bof{\frac{I_{n_{2}}\of{\beta}\e^{2\,\mu\,n_{2}}}{I_{n_{1}}\of{\beta}\e^{2\,\mu\,n_{1}}}}^{N_{t}}\\
=\e^{2\,N_{t}\,\sgn\of{n_{2}-n_{1}}\,\sum_{n'=\min\ssof{n_{1},n_{2}}}^{\max\ssof{n_{1},n_{2}}-1}\ssof{\mu-\mu_{n',n'+1}\of{\beta}}}\ ,\label{eq:o2weightratioz}
\end{multline}
which tell us, as function of the parameters $\mu$, $\beta$ and $N_{t}$, how strongly the term in the partition sum \eqref{eq:o2fluxreppartf1d} with $k=n_{2}$ is suppressed, compared to the term with $k=n_{1}$. Setting $n_{1}=0$ and $n_{2}=k$, the partition function \eqref{eq:o2fluxreppartf1d} can then be written as:
\[
Z=\ssof{I_{0}\ssof{\beta}}^{N_{t}}\,\sum\limits_{k=-\infty}^{\infty} W_{0,k}\of{\mu,\beta,N_{t}}\ .\label{eq:o2partfweightratios}
\]
If we now plug into expression \eqref{eq:o2weightratioz} for the weight ratios, that according to \eqref{eq:asymptoticbmucr} we have $\mu_{n,n+1}\ssof{\beta}\approx \ssof{2\,n+1}/\ssof{4\,\beta}$ for $\beta\gg 1$, expression \eqref{eq:o2weightratioz} simplifies to:
\[
W_{0,k}\of{\beta\,\mu,\kappa}\approx c\of{\beta\,\mu,\kappa} \e^{-\frac{\of{k-2\,\beta\,\mu}^{2}}{2\,\kappa}}\ ,\label{eq:o2weightratiolargebeta}
\]
with $c\ssof{\beta\,\mu,\kappa}=\e^{\frac{\of{2\,\beta\,\mu}^{2}}{2\,\kappa}}$, and the partition function \eqref{eq:o2partfweightratios} reduces to a sum over the values of a shifted Gaussian with variance $\kappa=\beta/N_{t}$ at integer values. If $\kappa$ is small, the Gaussian is strongly peaked around the $k$-value that is closest to $\ssof{2\,\beta\,\mu}$, so that there is always only one $k$-value that significantly contributes to the partition sum (unless $\beta\,\mu$ is fine-tuned to be exactly in the middle between two successive $k$-values). If $\ssof{2\,\beta\,\mu}$ changes from being closest to one $k$-value to being closest to the next $k$-value, the peak moves very abruptly from the old to the new dominant $k$, and the average $k$ changes in a step-like manner as function of $\beta\,\mu$. This is not the case if $\kappa$ is sufficiently large, as then the Gaussian is wide and also sub-dominant $k$-values contribute significantly to the partition sum, which allows the average $k$ value to change much more smoothly. The value of $\kappa$ at which the latter behavior can be expected to set in, can be approximated by the value for which the width of the $2\,\sigma$ interval of the Gaussian in \eqref{eq:o2weightratiolargebeta} does no longer fit between two successive $k$-values, i.e. if $2\times\ssof{2\,\sigma}\geq 1$, which is the case, if $\kappa=\sigma^{2}\geq 1/8$.\\
This can be made more explicitly by noting that the sum in \eqref{eq:o2partfweightratios} reduces for $\beta\gg 1$, using \eqref{eq:o2weightratiolargebeta}, to a \emph{Jacobi theta function}:
\[
\vartheta\of{u,q}=\sum\limits_{k=-\infty}^{\infty}u^{2\,k}\,q^{k^{2}}\ ,\label{eq:jacobifunc}
\]
with $q=\e^{-1/\of{2\,\kappa}}$ and $u=\e^{\beta\,\mu/\kappa}$. We can then use one of the \emph{Jacobi identities} for $\vartheta\of{u,q}$, namely:
\[
\vartheta\of{u'\of{u,q},q'\of{q}}=\alpha\of{u,q}\,\vartheta\of{u,q}\ ,
\]
with 
\begin{subequations}\label{eq:jacobiidentitytransf}
\[
q'\of{q}=\e^{\pi^{2}/\log\of{q}}=\e^{-2\,\pi^{2}\,\kappa}\ ,
\]
\[
u'\of{u,q}=\e^{\ii\,\pi\,\log\of{u}/\log\of{q}}=\e^{-\pi\,\ii\,\of{2\,\beta\,\mu}}\ ,
\]
and
\[
\alpha\of{u,q}=\sqrt{-\tfrac{\log\of{q}}{\pi}}\e^{\frac{\log^{2}\of{u}}{\log\of{q}}}=\frac{e^{-\frac{\of{2\,\beta\,\mu}^{2}}{2\,\kappa}}}{\sqrt{2\,\pi\,\kappa}}\ ,
\]
\end{subequations}
to write the partition function \eqref{eq:o2partfweightratios} as:
\begin{multline}
Z\overset{\beta\gg 1}{=}\of{I_{0}\of{\kappa\,N_{t}}}^{N_{t}}\vartheta\of{u,q}\\
=\of{I_{0}\of{\kappa\,N_{t}}}^{N_{t}}\vartheta\of{u'\of{u,q},q'\of{q}}/\alpha\of{u,q}\\
=\of{I_{0}\of{\kappa\,N_{t}}}^{N_{t}}\,\sqrt{2\,\pi\,\kappa}\,e^{\frac{\of{2\,\beta\,\mu}^{2}}{2\,\kappa}}\\
\cdot\vartheta\sof{\e^{-\pi\,\ii\,\of{2\,\beta\,\mu}},\e^{-2\,\pi^{2}\,\kappa}}\ .\label{eq:o2partflargebeta}
\end{multline}
For sufficiently large $\kappa$, the theta function on the last line converges towards unity as a small $q$ suppresses all terms with $k\neq 0$ in the defining sum \eqref{eq:jacobifunc}. The partition function then reduces to:
\[
Z\approx \of{I_{0}\of{\kappa\,N_{t}}}^{N_{t}}\,\sqrt{2\,\pi\,\kappa}\,e^{\frac{\of{2\,\beta\,\mu}^{2}}{2\,\kappa}}\ ,
\] 
and the corresponding charge density becomes:
\[
\avof{n}=\frac{1}{2\,N_{t}}\partd{\log\of{Z}}{\mu}=2\,\beta\,\mu\ ,
\]
with the linear behavior we observed in Fig.~\ref{fig:densvsmud1s0rescaledb}.\\

\subsection{Correlation function and mass spectrum}\label{ssec:corrfuncandmassspec}
Next, we would like to investigate how the two-point function of the (0+1)-dimensional $\On{2}$-model behaves as function of $\mu$. The general form of one- and two-point functions in the flux-variable formulation is derived in appendix~\ref{ssec:dualnpointfunc}. Here we will just need
\[
\avof{\phi^{-}_{x}\phi^{+}_{y}}\,=\,\frac{1}{Z}\spartd{Z}{s^{-}_{x}}{s^{+}_{y}}\,=\,\frac{Z_{2}^{-,+}\of{x,y}}{Z}\ ,\label{eq:o2twopointfuncgen}
\]
with $s^{\pm}_{x}$ being per-site sources\footnote{Note, that because of the continuum field $\phi\ssof{x}$ being dimensionless (the dimensionality is absorbed into $f_{\pi}$), the lattice fields $\phi^{\pm}_{x}$ do not explicitly scale with the lattice spacing $a$. The local \emph{per site sources}, $s_x^{\pm}$, do therefore also not explicitly scale with $a$; only the global $s^{\pm}$ do. This can be seen by noting that $\ssavof{\phi^{\pm}}=\frac{1}{V}\partd{Z}{s^{\pm}}$, while $\ssavof{\phi_{x}^{\pm}}=\partd{Z}{s^{\pm}_{x}}$, where $\ssavof{\phi^{\pm}}$ and $\ssavof{\phi^{\pm}_{x}}$ are both dimensionless and in the case of translation invariance, should even be equal, $\ssavof{\phi^{\pm}}=\ssavof{\phi^{\pm}_{x}}$, $\forall x$.}. For the partition function \eqref{eq:o2fluxreppartf1d} (or equivalently \eqref{eq:o2partfweightratios}), this two-point function takes the simple form:
\begin{multline}
\avof{\phi^{-}_{x}\phi^{+}_{y}}\of{\mu,\beta,N_{t}}=\\
\frac{\sum\limits_{k=-\infty}^{\infty}W_{0,k}^{1-s\of{x,y}}\of{\mu,\beta,N_{t}}\,W_{0,k+1}^{s\of{x,y}}\of{\mu,\beta,N_{t}}}{2\,\sum\limits_{k=-\infty}^{\infty}W_{0,k}\of{\mu,\beta,N_{t}}}\ ,\label{eq:o2fluxreptwopointfunc1d}
\end{multline}
with $s\ssof{x,y}=\of{\of{y-x}\bmod N_{t}}/N_{t}$, and where $\of{X\bmod Y}$ with $X\in\mathbb{Z}$ and $Y\in\mathbb{N}$ is the modulo operation, that wraps $X$ around the interval $\cof{0,\ldots,Y-1}$. The $W_{n_{1},n_{2}}\ssof{\mu,\beta,N_{t}}$ are the weight ratios introduced in \eqref{eq:o2weightratioz}. In order to avoid too lengthy expression we will often leave the dependency of quantities on the parameters $\mu$, $\beta$ and $N_{t}$ implicit in the following. Equation \eqref{eq:o2fluxreptwopointfunc1d} simply states that if a $\phi^{-}$ is inserted at site $x$ and a $\phi^{+}$ at site $y$, then, if the source $\tilde{s}$ is set to zero, $\tilde{s}=0$, the two sites have to be connected by a continuous charged flux-line.\\
For $s\ssof{x,y}<1/2$, it is useful to write \eqref{eq:o2fluxreptwopointfunc1d} in the form:
\begin{subequations}\label{eq:o2twopointfnc}
\begin{multline}
\avof{\phi^{-}_{x}\phi^{+}_{y}}=\frac{\sum\limits_{k=-\infty}^{\infty}W_{0,k}\,W_{k,k+1}^{s\of{x,y}}}{2\,\sum\limits_{k=-\infty}^{\infty}W_{0,k}}\\
=\frac{\sum\limits_{k=-\infty}^{\infty}W_{0,k}\,\e^{2\,N_{t}\,\of{\mu-\mu_{k,k+1}}\,s\of{x,y}}}{2\,\sum\limits_{k=-\infty}^{\infty}W_{0,k}}\ ,\label{eq:o2twopointfncfwd}
\end{multline}
where on the second line, we made use of the last line of \eqref{eq:o2weightratioz} to write the $W_{k,k+1}$ in terms of exponentials, from which one can easily read off the masses of all the propagating modes.
The restriction, $s\ssof{x,y}<1/2$, ensures that the largest term in the sum in the numerator of \eqref{eq:o2twopointfncfwd} is the one with $k=n$ if $\mu\in\loint{\mu_{n-1,n}\of{\beta},\mu_{n,n+1}\of{\beta}}$.
Similarly, one can for $s\ssof{x,y}>1/2$ write:
\begin{multline}
\avof{\phi^{-}_{x}\phi^{+}_{y}}=\frac{\sum\limits_{k=-\infty}^{\infty}W_{0,k}\,W_{k,k-1}^{1-s\of{x,y}}}{2\,\sum\limits_{k=-\infty}^{\infty}W_{0,k}}\\
=\frac{\sum\limits_{k=-\infty}^{\infty}W_{0,k}\,\e^{-2\,N_{t}\,\of{\mu-\mu_{k-1,k}}\,s\of{y,x}}}{2\,\sum\limits_{k=-\infty}^{\infty}W_{0,k}}\ ,\label{eq:o2twopointfncbwd}
\end{multline}
\end{subequations}
where the term with $k=n$ in the sum in the numerator of \eqref{eq:o2twopointfncbwd} is this time largest if $\mu\in\roint{\mu_{n-1,n}\of{\beta},\mu_{n,n+1}\of{\beta}}$.\\
From eqns. \eqref{eq:o2twopointfncfwd} and \eqref{eq:o2twopointfncbwd} we can see that the spectrum contains a massless mode whenever the chemical potential $\mu$ is set to one of the critical values $\mu_{n,n+1}$, $n\in\mathbb{Z}$.\\
\begin{figure}[ht]
\centering
\begin{minipage}[t]{0.48\textwidth}
\centering
\includegraphics[width=0.85\linewidth]{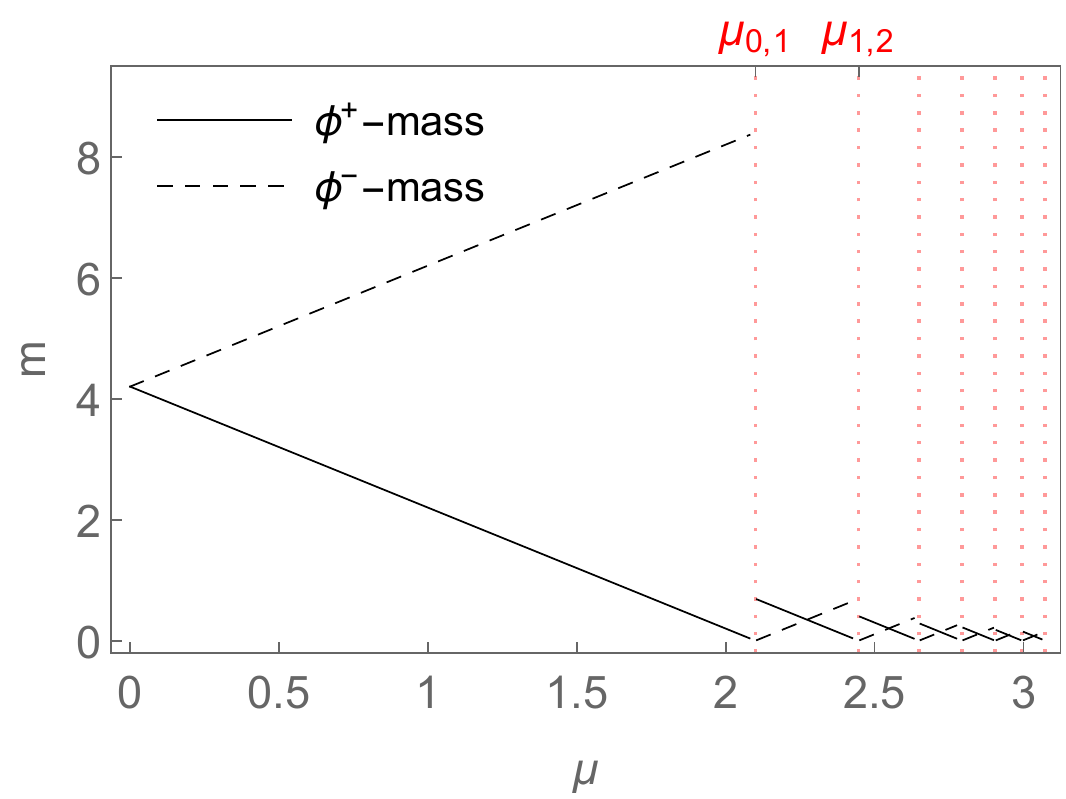}
\caption{The figure shows for the one-dimensional $\On{2}$-model with $\tilde{s}=0$ and $\beta=0.03$ the zero-temperature $\phi^{\pm}$-masses from eqn. \eqref{eq:o2phiplusmass1dgenmu} and \eqref{eq:o2phiminusmass1dgenmu} as functions of the chemical potential $\mu$.}
\label{fig:massesvsmu1ds0}
\end{minipage}
\end{figure}

At fixed $\beta$ and in the limit $\ssof{N_{t}\to\infty}$, the $\phi^{\pm}$-masses can also be extracted in the standard way from the exponential decay of the connected two-point function. To this end, we first note, that for $\tilde{s}=0$ the $\Un{1}$-current is conserved and the $\phi^{\pm}$-worldlines, which carry a $\Un{1}$-charge, therefore need to be continuous. This means that $\ssavof{\phi_{x}^{-}\phi_{y}^{+}}$ is already a \emph{connected two-point function} whose large distance behavior is dominated by the lightest mode and not by any disconnected piece. For $\mu\in\loint{\mu_{n-1,n}\ssof{\beta},\mu_{n,n+1}\ssof{\beta}}$, the $\phi^{+}$-mass is then obtained from \eqref{eq:o2twopointfncfwd} with $1\ll t\ll t_{*}$ by computing:
\begin{subequations}\label{o2phiplusminusmass1dgenmu}
\begin{multline}
m^{+}\of{\mu,\beta}=\lim\limits_{\mathclap{N_{t}\to\infty}}\,\log\bof{\frac{\avof{\phi^{-}_{0}\phi^{+}_{t}}}{\avof{\phi^{-}_{0}\phi^{+}_{t+1}}}}\\
=-\lim\limits_{\mathclap{N_{t}\to\infty}}\,\log\of{W_{n,n+1}}/N_{t}\\
=\,2\,\of{\mu_{n,n+1}-\mu}\ ,\label{eq:o2phiplusmass1dgenmu}
\end{multline}
where $t_{*}$ is given by the value of $t$ for which the amplitude of the lightest forward-propagating mode becomes smaller than the amplitude of the lightest backward-propagating mode (if $\mu=0$, then $t_{*}=N_{t}/2$).
Similarly, using that $\ssavof{\phi^{+}_{0}\phi^{-}_{t'}}=\ssavof{\phi^{-}_{0}\phi^{+}_{N_{t}-t'}}$, the $\phi^{-}$-mass can for $\mu\in\roint{\mu_{n-1,n}\ssof{\beta},\mu_{n,n+1}\ssof{\beta}}$ be obtained from \eqref{eq:o2twopointfncbwd} with $1\ll t'\ll N_{t}-t_{*}$ as:
\begin{multline}
m^{-}\of{\mu,\beta}=\lim\limits_{\mathclap{N_{t}\to\infty}}\,\log\bof{\frac{\avof{\phi^{-}_{0}\phi^{+}_{N_{t}-t'}}}{\avof{\phi^{-}_{0}\phi^{+}_{N_{t}-t'-1}}}}\\
=-\lim\limits_{\mathclap{N_{t}\to\infty}}\,\log\of{W_{n-1,n}}/N_{t}\\
=\,2\,\of{\mu-\mu_{n-1,n}}\ ,\label{eq:o2phiminusmass1dgenmu}
\end{multline}
\end{subequations}
with $t'$ kept fixed while $N_{t}$ is sent to infinity.\\
The masses \eqref{eq:o2phiplusmass1dgenmu} and \eqref{eq:o2phiminusmass1dgenmu} are shown in Fig.~\ref{fig:massesvsmu1ds0} for $\beta=0.03$.\\
With the expressions for the $\phi^{\pm}$-masses, we can now also write down an expression for $t_{*}$: requiring that $\e^{-m^{+}\ssof{\mu,\beta}\,t_{*}}=\e^{-m^{-}\ssof{\mu,\beta}\,\ssof{N_{t}-t_{*}}}$ and solving this for $t_{*}$, we find:
\begin{multline}
t_{*}\of{\mu,\beta,N_{t}}=\frac{m^{-}\of{\mu,\beta}\,N_{t}}{m^{+}\of{\mu,\beta}-m^{-}\of{\mu,\beta}}\\
=\frac{\of{\mu-\mu_{n-1,n}}\,N_{t}}{\mu_{n,n+1}-\mu_{n-1,n}}\ ,\label{eq:tstar}
\end{multline}
where the $n$ is such that $\mu\in\ssof{\mu_{n-1,n},\mu_{n,n+1}}$.\\

As the Euclidean two-point function \eqref{eq:o2twopointfnc} is a sum of pure exponentials $\sim \e^{-m\,t}$, a zero mode implies the presence of a constant piece in the two-point function. In the vicinity of a critical point $\mu_{n,n+1}$ and for sufficiently large $N_{t}$, this constant piece can be obtained from the value of the two-point function \eqref{eq:o2fluxreptwopointfunc1d} at $t=t_{*}$ with $t_{*}$ from \eqref{eq:tstar}. For $\mu$ slightly below $\mu_{n,n+1}$, one finds:
\begin{subequations}\label{eq:o2fluxrepcorrdisconnpiece1d}
\begin{multline}
\savof{\phi^{-}_{0}\phi^{+}_{t_{*}}}\approx\frac{W_{0,n}\,W_{n,n+1}^{s_{*}}+\ldots}{2\,\of{W_{0,n}+W_{0,n+1}+\ldots}}\\
\approx\frac{\,\e^{2\,N_{t}\frac{\of{\mu-\mu_{n,n+1}}\of{\mu-\mu_{n-1,n}}}{\mu_{n,n+1}-\mu_{n-1,n}}}}{2\,\of{1+\e^{2\,N_{t}\,\of{\mu-\mu_{n,n+1}}}}}\ , \label{eq:o2fluxrepcorrdisconnpiece1da}
\end{multline}
and for $\mu$ slightly above $\mu_{n,n+1}$, correspondingly:
\begin{multline}
\savof{\phi^{-}_{0}\phi^{+}_{t_{*}}}\approx\frac{W_{0,n+1}\,W_{n+1,n+2}^{s_{*}}+\ldots}{2\,\of{W_{0,n}+W_{0,n+1}+\ldots}}\\
\approx\frac{\e^{2\,N_{t}\frac{\of{\mu-\mu_{n+1,n+2}}\of{\mu-\mu_{n,n+1}}}{\mu_{n+1,n+2}-\mu_{n,n+1}}}}{2\,\of{1+\e^{-2\,N_{t}\,\of{\mu-\mu_{n,n+1}}}}}\ , \label{eq:o2fluxrepcorrdisconnpiece1db}
\end{multline}
\end{subequations}
where $s_{*}=t_{*}\of{\mu,\beta,N_{t}}/N_{t}$ and we made again used of \eqref{eq:o2weightratioz}. For sufficiently large $N_{t}$, \eqref{eq:o2fluxrepcorrdisconnpiece1d} assumes at a critical point $\mu=\mu_{n,n+1}$, $n\in\mathbb{Z}$ the value $1/4$ and decays rapidly as $\mu$ moves away from $\mu_{n,n+1}$.\\

So far we have only looked at the behavior of the two-point function at fixed $\beta$, in which case, as we have seen in the previous section, the limit $\ssof{N_{t}\to\infty}$ gives rise to the appearance of first-order transitions at the very same values of the chemical potential, for which the zero modes and the constant piece appear in the two-point function (c.f. eq. \eqref{eq:o2fluxrepcorrdisconnpiece1d}). Do these zero modes and constant pieces survive when the limit $\ssof{N_{t}\to\infty}$ is taken at fixed $\kappa=\beta/N_{t}=T\,f_{\pi}^{2}$ instead of fixed $\beta$, in which case there are no first order transitions?\\

As we have seen already in the previous section, the behavior of the system as function of the chemical potential changes when we go from low to sufficiently high temperatures. This should also be reflected in a change of behavior in the two-point function. To analyse the latter, we use again expression \eqref{eq:o2weightratiolargebeta} for the weight ratios $W_{n_1,n_2}\of{\beta\,\mu,\kappa}$ at large $\beta$ and write the two-point function as:
\begin{multline}
\avof{\phi_{x}^{-}\phi_{y}^{+}}\of{\beta\,\mu,\kappa}=\frac{\sum\limits_{k=-\infty}^{\infty}W_{0,k}\,W_{k,k+1}^{s\of{x,y}}}{2\,\sum\limits_{k=-\infty}^{\infty}W_{0,k}}\\
\overset{\beta\gg 1}{=}\frac{\sum\limits_{k=-\infty}^{\infty}\e^{-\frac{\of{k-2\,\beta\,\mu}^{2}}{2\,\kappa}}\,\e^{-\frac{\of{2\,k+1-4\,\beta\,\mu}\,s\of{x,y}}{2\,\kappa}}}{2\,\sum\limits_{k=-\infty}^{\infty}\e^{-\frac{\of{k-2\,\beta\,\mu}^{2}}{2\,\kappa}}}\\
=\frac{\sum\limits_{k=-\infty}^{\infty}\e^{-\frac{\of{k-2\,\beta\,\mu}^{2}}{2\,\kappa}}\,\e^{-\frac{\beta\,m_{k}\,s\of{x,y}}{\kappa}}}{2\,\sum\limits_{k=-\infty}^{\infty}\e^{-\frac{\of{k-2\,\beta\,\mu}^{2}}{2\,\kappa}}}\ .\label{eq:o2twopointfunclargebeta}
\end{multline}
with mass terms 
\[
\beta\,m_{k}\of{\beta\,\mu,\kappa}=\frac{2\,k+1-4\,\beta\,\mu}{2}\ \forall\,k\in\mathbb{Z}\ .
\]

As long as the temperature is sufficiently small, i.e. for $\kappa=\beta/N_{t}=f_{\pi}\,T<1/8$, the Gaussian weights in numerator and denominator of \eqref{eq:o2twopointfunclargebeta} are sufficiently narrow to give dominant weight to the terms for which $k$ is closest to the value $2\,\beta\,\mu$.
At a critical point
\[
\beta\,\mu=\beta\,\mu_{n,n+1}\overset{\beta\gg 1}{=}\ssof{2\,n+1}/4\ ,\ n\in\mathbb{Z}\ ,
\]
the value of $2\,\beta\,\mu$ is located precisely in the middle between two successive $k$-values. The Gaussian weights for the terms with $k=n$ and $k=n+1$ are then equal, and the masses corresponding to these two dominant modes assume the values $\beta\,m_n=0$ and $\beta,m_{n+1}=1$, meaning that the long-distance behavior of the system is dominated by a zero mode.\\

\begin{figure}[ht]
\centering
\begin{minipage}[t]{0.49\textwidth}
\centering
\includegraphics[width=0.85\linewidth]{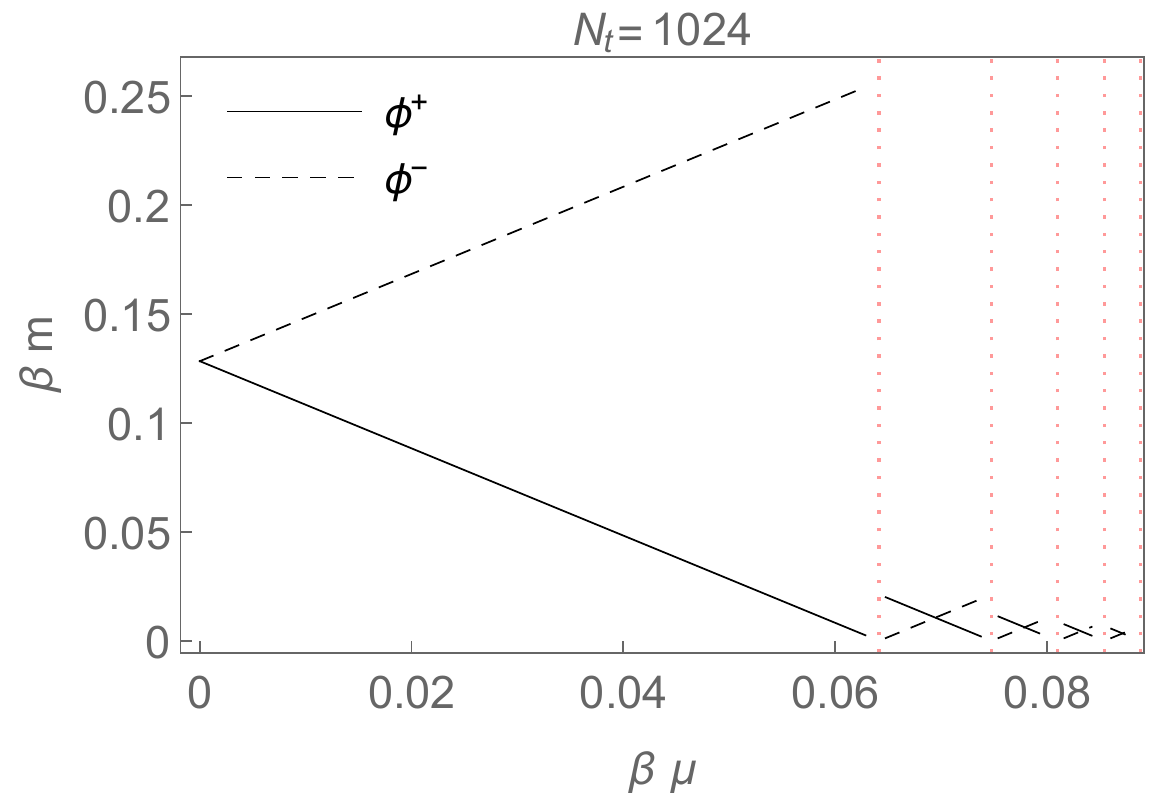}
\end{minipage}\\[5pt]
\begin{minipage}[t]{0.49\textwidth}
\centering
\includegraphics[width=0.85\linewidth]{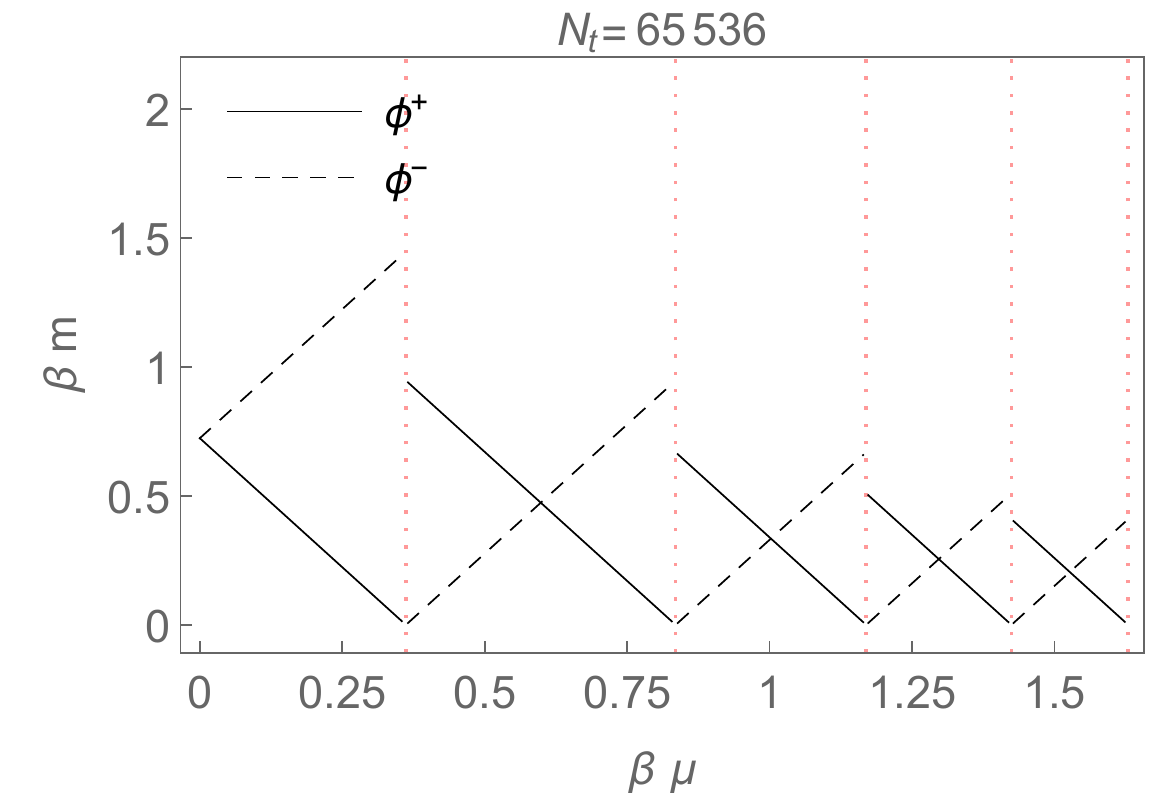}
\end{minipage}\\[5pt]
\begin{minipage}[t]{0.49\textwidth}
\centering
\includegraphics[width=0.85\linewidth]{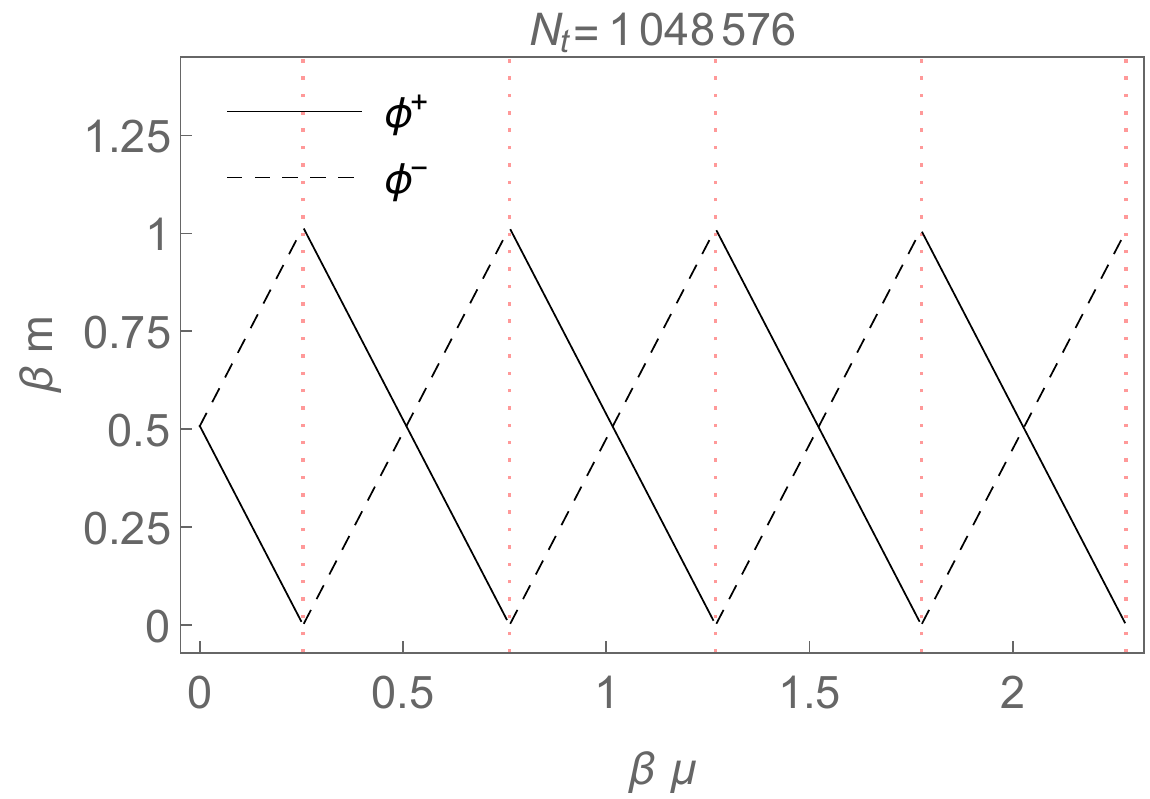}
\end{minipage}
\caption{The figures show for the one-dimensional $\On{2}$-model with $\tilde{s}=0$ and $\kappa=\beta/N_{t}=0.00003$ the rescaled $\phi^{\pm}$-masses, $\beta\,m^{\pm}$, as functions of $\beta\,\mu$ for different values of $N_{t}$.}
\label{fig:hmassesvshmu1ds0}
\end{figure}

In Fig.~\ref{fig:hmassesvshmu1ds0} the $N_{t}$-dependency of the rescaled masses $\beta\,m^{\pm}$ as functions of $\beta\,\mu$ is illustrated for $\kappa=0.00003$, corresponding to a low but non-zero temperature. The spectrum in the uppermost panel with $N_{t}=1024$ then corresponds to a system at $\beta=\kappa\,N_{t} \approx 0.03$ and looks therefore (up to the rescaling of the axes by a factor of $\beta$) identical to the spectrum shown in Fig.~\ref{fig:massesvsmu1ds0}. However, if now $N_{t}$ is increased while $\kappa$ is kept fixed, $\beta$ has to increases too and the critical values of $\beta\,\mu$ change according to \eqref{eq:asymptoticbmucr}: as long as $\beta=\kappa\,N_{t}\ll 1$, the critical values of $\beta\,\mu$ grow like $\beta\,\mu_{n,n+1}\ssof{\beta}\approx \frac{\beta}{2}\log\sof{\frac{2\ssof{n+1}}{\beta}}$ as function of increasing $\beta=\kappa\,N_{t}$, but as soon as $N_{t}$ is sufficiently large, so that $\beta=\kappa\,N_{t}\gg 1$, the critical values of $\beta\,\mu$ converge towards their $\beta$-independent asymptotic values $\beta\,\mu_{n,n+1}\ssof{\beta}\approx \frac{2\,n+1}{4}$, $n\in\mathbb{Z}$. The critical values then become uniformly spaced an the mass-spectrum, which depends on the difference between successive critical values, becomes periodic in $\beta\,\mu$, as illustrated in the lower-most panel of Fig.~\ref{fig:hmassesvshmu1ds0}. At this point, a further increase of $N_{t}$ leaves the spectrum unchanged.\\

For $\kappa$ bigger than $1/8$, the Gaussian weights in \eqref{eq:o2twopointfunclargebeta} can no longer sufficiently project on the terms for which $k$ is closest to $2\,\beta\,\mu$ and the long-distance behavior of the system at a critical value of $\beta\mu$ will not necessarily be dominated by a zero mode any more. By expressing the two-point function in terms of theta functions, we can write:
\begin{multline}
\avof{\phi_{x}^{-}\phi_{y}^{+}}\of{\beta\,\mu,\kappa}\overset{\beta\gg 1}{=}\frac{\of{u^{2}\,q}^{s\of{x,y}}\vartheta\of{u_{2}\of{x,y},q}}{2\,\vartheta\of{u,q}}\\
=\frac{\of{u^{2}\,q}^{s\of{x,y}}\alpha\of{u,q}\,\vartheta\of{u'\of{u_{2}\of{x,y},q},q'\of{q}}}{2\,\alpha\of{u_{2}\of{x,y},q}\,\vartheta\of{u'\of{u,q},q'\of{q}}}\\
=\e^{-\frac{s\of{x,y}\,\of{1-s\of{x,y}}}{2\,\kappa}}\\
\cdot\frac{\vartheta\of{\e^{-\pi\,\ii\,\of{2\,\beta\,\mu-s\of{x,y}}},\e^{-2\,\pi^{2}\,\kappa}}}{2\,\vartheta\of{\e^{-\pi\,\ii\,\of{2\,\beta\,\mu}},\e^{-2\,\pi^{2}\,\kappa}}}\ ,\label{eq:o2twopointfunclargebetatheta}
\end{multline}
where $q=\e^{-1/\of{2\,\kappa}}$, $u=\e^{\beta\,\mu/\kappa}$, $u_{2}\of{x,y}=u\,q^{s\of{x,y}}$ and $s\ssof{x,y}=\of{\of{y-x}\bmod N_{t}}/N_{t}$, and on the second line, the transformations \eqref{eq:jacobiidentitytransf} were used. For sufficiently large $\kappa$, the theta functions on the last line of \eqref{eq:o2twopointfunclargebetatheta} converge towards unity and the two-point function reduces therefore to:
\[
\avof{\phi_{x}^{-}\phi_{y}^{+}}\of{\kappa}\approx\frac{1}{2}\,\e^{-\frac{s\of{x,y}\,\of{1-s\of{x,y}}}{2\,\kappa}}\ ,
\]
which is independent of $\beta\,\mu$ but depends only on the temperature $\sim \kappa$.\\
To summarize: the expression \eqref{eq:o2twopointfunclargebeta} for the two-point function illustrates that zero modes will always be present in the large-$N_{t}$ limit when $\mu$ is set to one of the critical points $\mu_{n,n+1}$, $n\in\mathbb{Z}$, regardless of whether $\kappa$ is small or large. However, if $\kappa$, which sets the temperature, is sufficiently large, the system will be completely dominated by thermal modes and the zero-modes will no-longer give relevant contributions to expectation values at the critical points. 

\subsection{Explicit symmetry breaking}\label{ssec:explicitsymmbreak}
We have seen in the previous section that the $\On{2}$ model, coupled to a chemical potential $\mu$, can in (0+1) dimensions develop some sort of long-range order when $\mu$ approaches certain critical values $\mu_{n,n+1}$, $n\in\mathbb{Z}$. In this section we would now like to check whether this long range order also gives rise to the formation of a non-zero condensate $\ssavof{\phi^{\pm}}$.\\

To measure a condensate directly, one needs to break the global $\Un{1}$ symmetry explicitly by setting $\tilde{s}$ in \eqref{eq:o2fluxreppartfg} to a non-zero value. This is necessary because of the fact, that the partition function \eqref{eq:o2fluxreppartfg}, although reformulated in terms of flux-variables, still represents a path-integral that sums over all possible configurations of the lattice field $\phi^{\pm}_{x}$, which means, that if the global symmetry is spontaneously broken, the path-integral still sums over all possible ways in which this spontaneous breaking can happen. In expectation values of observables like $\phi^{\pm}=\frac{1}{\sqrt{2}}\e^{\pm\,\ii\,\theta}$, which depend on the direction in which the symmetry gets broken, the contributions from configurations that differ only by this direction would therefore exactly cancel each other. To avoid such cancellations, one has to single out a particular direction in which the symmetry should break, which is most easily done by coupling $\phi^{\pm}_{x}$ to a non-zero source $s^{\pm}=\frac{\tilde{s}\,\e^{\mp\,\ii\,\phi_{s}}}{\sqrt{2}}$. The value of the partition function does not depend on the choice of the direction that is singled out and the choice of the phase $\phi_{s}$ of the source is therefore irrelevant; only the magnitude $\tilde{s}$ of the source matters. The value of $\phi_{s}$ will only enter in expectation values of observables that are not invariant under the global $\Un{1}$ symmetry.\\

With $\tilde{s}>0$, the 1D partition function for the $\On{2}$-model does not reduce to a single sum as in equation \eqref{eq:o2fluxreppartf1d}; the non-zero value of $\tilde{s}$ breaks charge conservation, so that $k$-variables on different links can now take on different values, at the cost of producing monomers and thereby changing the site weight $I_{p_{x}}\ssof{\tilde{s}}$. One then arrives at the following expression:
\begin{multline}
Z=\sum\limits_{\cof{k}}\,\prod\limits_{x}\bcof{I_{k_{x}}\of{\beta}\,e^{2\,\mu\,k_{x}}\\
\cdot I_{\of{k_{x}-k_{x-1}}}\of{\tilde{s}}\,\e^{\ii\,\phi_{s}\,\of{k_{x}-k_{x-1}}}}\\
=\,\trace\sof{T_{0}^{N_{t}}}\ ,\label{eq:o2fluxreppartf1dincs}
\end{multline}
where we have summed over the $p$-variables in order to get rid of the delta-function constraints in \eqref{eq:o2fluxreppartfg}, and after the second equality sign, $N_{t}$ is (as usual) the temporal system size. The \emph{transfer matrix}, $T_{\delta}$, can for example be chosen as:
\begin{multline}
\of{T_{\delta}}_{a b}=\\
I_{a}\of{\beta}\,\e^{2\,\mu\,a}\,\frac{I_{\of{a-b-\delta}}\of{\tilde{s}}\,\e^{\ii\,\phi_{s}\,\of{a-b-\delta}}}{2^{\abs{\delta}/2}}\ ,\label{eq:o2fluxrep1dtransfermat}
\end{multline}
with $\delta\in\cof{-1,1,0}$, depending on whether the site to which the transfer matrix is associated contains an external $\phi^{-}$, $\phi^{+}$ or no external field (cf. the discussion in appendix~\ref{ssec:dualnpointfunc} and the appearance of Kronecker-deltas in the delta-function constraints for the one- and two-point partition functions).\\
The one- and two-point functions can be expressed in terms of the transfer matrix as:
\[
\avof{\phi^{\pm}}\,=\,\frac{\trace\sof{T_{\pm 1}^{}\,T_{0}^{N_{t}-1}}}{\trace\sof{T_{0}^{N_{t}}}}\ ,\label{eq:o2cond1dfinites}
\]
and
\begin{multline}
\avof{\phi^{-}_{0}\phi^{+}_{t}}=\\
\ucases{\frac{1}{2}&\text{if}\quad t=0\\\frac{\trace\of{T_{-1}^{}T_{0}^{t-1}T_{1}^{\vphantom{N_{t}}}T_{0}^{N_{t}-t-1}}}{\trace\of{T_{0}^{N_{t}}}}&\text{if}\quad t>0}\ .\label{eq:o2corr1dfinites}
\end{multline}
\begin{figure}[ht]
\centering
\begin{minipage}[t]{0.49\textwidth}
\centering
\includegraphics[width=0.85\linewidth]{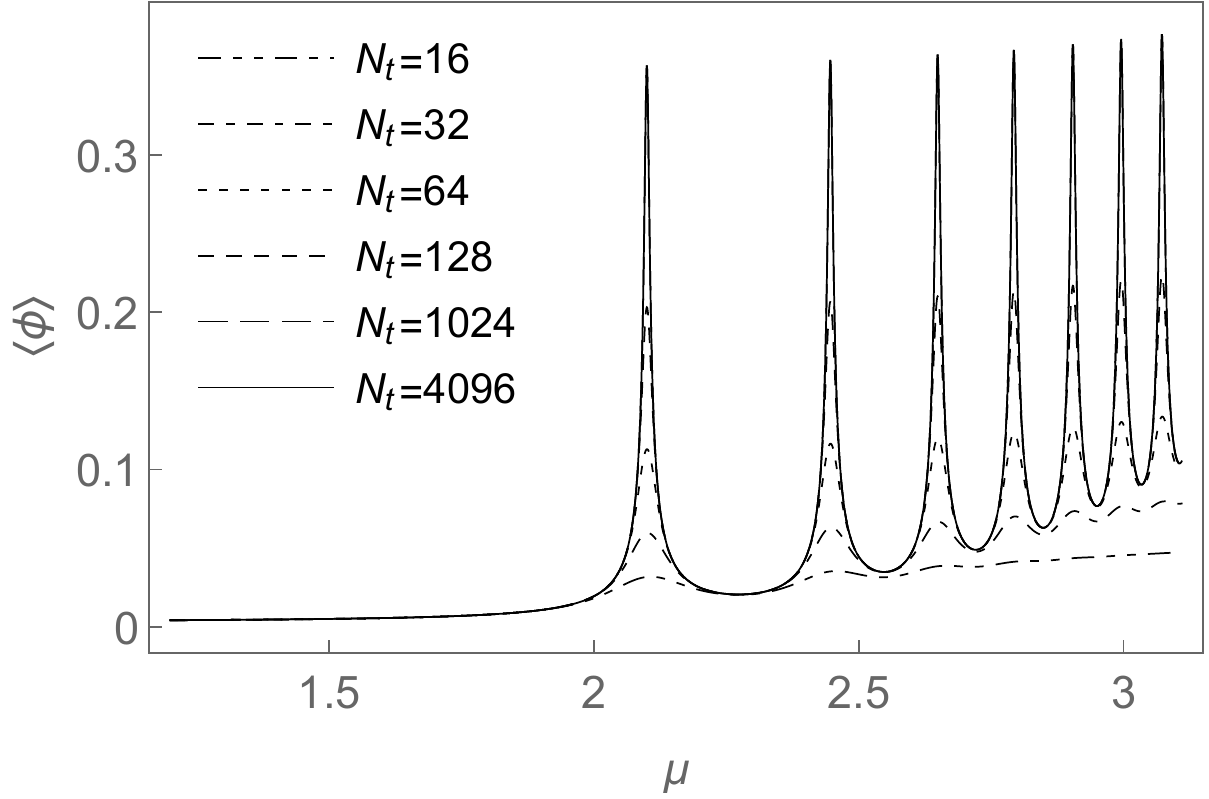}
\caption{The figure shows the $\phi^{\pm}$-condensate as function of the chemical potential $\mu$ for the (0+1)-dimensional $\On{2}$-model with $\tilde{s}=0.01$ and $\beta=0.03$, obtained from eq. \eqref{eq:o2cond1dfinites} for various values of $N_{t}$. The curves for $N_{t}=1024$ and $N_{t}=4096$ are essentially on top of each other.}
\label{fig:o2condvsmu1d}
\end{minipage}
\end{figure}
In Fig.~\ref{fig:o2condvsmu1d}, we show \eqref{eq:o2cond1dfinites} as function of $\mu$ when evaluated numerically with $\tilde{s}=0.01$ and $\beta=0.03$ for various values of $N_{t}$. As can be seen: at the critical points, $\mu=\mu_{n,n+1}$, $n\in\mathbb{Z}$, the condensate, $\ssavof{\phi^{\pm}}$, peaks and approaches for sufficiently large $N_{t}$ a finite value. A comparison of the value of the condensate and the magnitude of the constant piece of the two-point function \eqref{eq:o2corr1dfinites} at these critical points further shows, that:
\[
\lim\limits_{t\to\infty}\avof{\phi^{-}_{0}\phi^{+}_{t}}\,=\,\avof{\phi^{-}}\avof{\phi^{+}}\ .\label{eq:asymptcorrcondrel}
\]
However, comparing the numerically obtained result for this constant piece in the two point function for $\tilde{s}>0$ with the value we found above in \eqref{eq:o2fluxrepcorrdisconnpiece1d} for $\tilde{s}=0$, reveals, that the value for $\tilde{s}>0$ is only about half the value obtained with $\tilde{s}=0$.\\

\begin{figure*}[ht]
\centering
\begin{minipage}[t]{0.49\linewidth}
\centering
\includegraphics[width=0.9\linewidth]{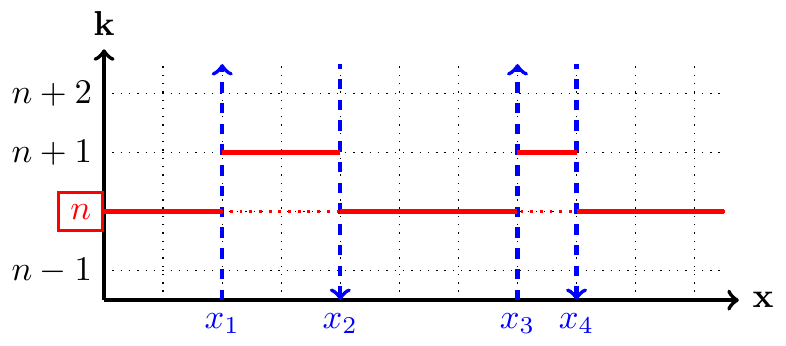}
\end{minipage}\hfill
\begin{minipage}[t]{0.49\linewidth}
\centering
\includegraphics[width=0.9\linewidth]{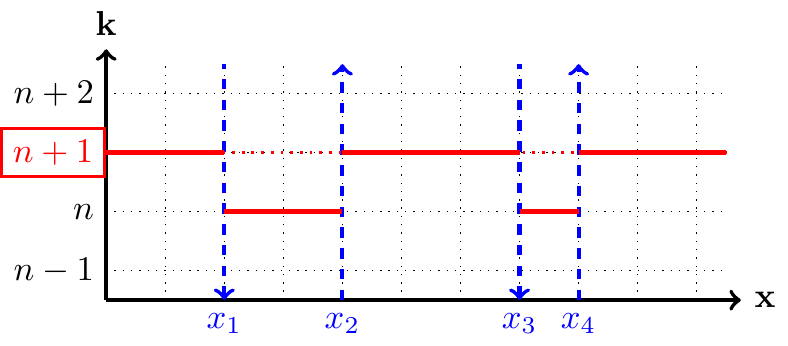}
\end{minipage}
\caption{The figure shows how the function \eqref{eq:binarymodelpartf} enters the denominator of \eqref{eq:o2corr1dfinitesapprox}. The $x$-axis represents the different sites of the periodic (0+1)-dimensional lattice on which \eqref{eq:o2fluxreppartf1dincs} is defined, and the $y$-axis display the values of the $k$-variables between the sites. The chemical potential is set to the critical value $\mu=\mu_{n,n+1}$, so that $k_{x}=n$ $\forall x$ and $k_{x}=n+1$ $\forall x$ are equally likely and have maximum weight. However, as $\tilde{s}>0$, the $k$-variables don't have to be all the same, but can vary between the two values $n$ and $n+1$, while other values are highly suppressed. Now, assume that we are initially in the configuration where $k_{x}=n$ $\forall x$ (left-hand panel). As all $k$-variables are equal, all site-weights in \eqref{eq:o2fluxreppartf1dincs} assume the value $I_{0}\of{\tilde{s}}$. Next, we modify this configuration by picking some \emph{shift-points}, e.g. $\cof{x_{1},\ldots,x_{4}}$, at which the values of incoming and outgoing $k$-variable changes form $n$ to $n+1$ (upward shift-point) or from $n+1$ back to $n$ (downward shift-point). As the lattice is periodic, these shift-points have to appear in pairs. The modified configuration has lower weight than the original one, as the site-weights for the shift-points $\cof{x_{1},\ldots,x_{4}}$ in \eqref{eq:o2fluxreppartf1dincs} are reduced from $I_{0}\of{\tilde{s}}$ to $I_{1}\of{\tilde{s}}$; but, there are $\binom{N_{t}}{4}$ ways of picking the four shift-points, $\cof{x_{1},\ldots,x_{4}}$, and therefore $\binom{N_{t}}{4}$ such modified configurations which are distinct but have the same weight. Similarly, if the number of shit-points is not four, but more generally $2\,m$, with $m\in\cof{0,1,\ldots,\floor{N_{t}/2}}$, there will be $\binom{N_{t}}{2\,m}$ configurations with a weight in which $2\,m$ site-weights have value $I_{1}\of{\tilde{s}}$ instead of $I_{0}\of{\tilde{s}}$. The total weight of all the modified configurations that can be obtained in this way from the config. with $k_{x}=n$ $\forall x$ (including the config. with $k_{x}=n$ $\forall x$ itself) is therefore: $\sum_{m=0}^{\floor{N_{t}/2}}\,\binom{N_{t}}{2\,m}\,I_{0}^{N_{t}-2\,m}\of{\tilde{s}}\,I_{1}^{2\,m}\of{\tilde{s}}$. In the same way, one can modify the configuration with $k_{x}=n+1$ $\forall x$ (right-hand panel). One then gets the set of configurations which are the mirror-images of the ones obtained from the config. with $k_{x}=n$ $\forall x$, (i.e. the values $n$ and $n+1$ of the $k$-variables are interchanged), but as they are equally many mirror-images, the total weight is again given by: $\sum_{m=0}^{\floor{N_{t}/2}}\,\binom{N_{t}}{2\,m}\,I_{0}^{N_{t}-2\,m}\of{\tilde{s}}\,I_{1}^{2\,m}\of{\tilde{s}}$.}
\label{fig:o2partf1dmonomerinsertion}
\end{figure*}

\begin{figure*}[ht]
\centering
\begin{minipage}[t]{0.49\linewidth}
\centering
\includegraphics[width=0.9\linewidth]{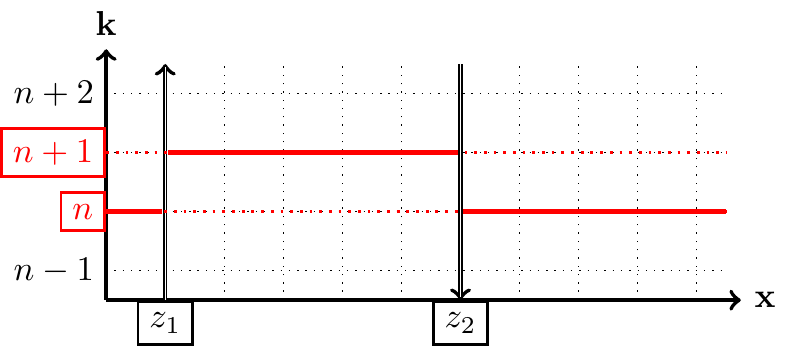}
\end{minipage}\hfill
\begin{minipage}[t]{0.49\linewidth}
\centering
\includegraphics[width=0.9\linewidth]{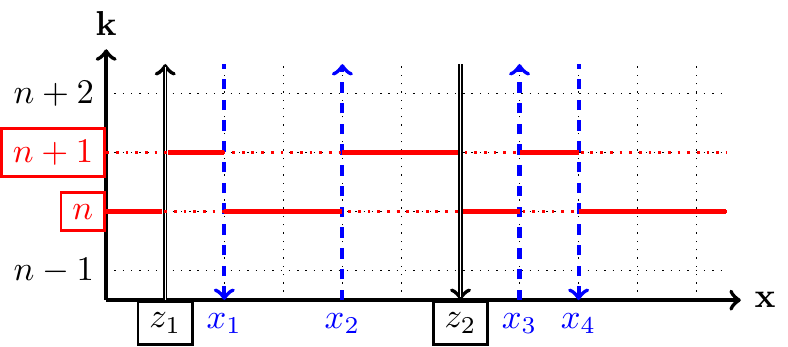}
\end{minipage}
\caption{The figure illustrates how the function \eqref{eq:binarymodelpartf} enters the enumerator of \eqref{eq:o2corr1dfinitesapprox} and we proceed in a similar way as in Fig.~\ref{fig:o2partf1dmonomerinsertion}. The chemical potential is set to the critical value $\mu=\mu_{n,n+1}$, so that all $k$-variables have value $n$ or $n+1$, and we start with the maximum-weight configuration, displayed in left-hand panel, in which the external $\phi^{-}$ at site $z_{1}$ requires the incoming and outgoing $k$-variables to have values $n$ and $n+1$, respectively, while the $\phi^{+}$ at site $z_{2}$ forces incoming and outgoing $k$-variables to have values $n+1$ and $n$. As in Fig.~\ref{fig:o2partf1dmonomerinsertion}, we can now pick an even number of lattice sites which become \emph{shift-points} at which the values of incoming and outgoing $k$-variables have to be $n$ and $n+1$, respectively (upward shift-point), or $n+1$ and $n$ (downward shift-point), unless the shift-point coincides with $z_{1}$ or $z_{2}$, in which case incoming and outgoing $k$-variable have to have the the same value instead of distinct ones. In the right-hand panel, we have again choose four such shift-points, $\cof{x_{1},\ldots,x_{4}}$, and obtain a modified configuration for which four site-weights are changed form $I_{0}\of{\tilde{s}}$ to $I_{1}\of{\tilde{s}}$. But, in contrast to the situation in Fig.~\ref{fig:o2partf1dmonomerinsertion}, the presence of the external fields on sites $z_{1}$ and $z_{2}$, limits the number of ways in which one can pick the four shift-points, $\cof{x_{1},\ldots,x_{4}}$: $z_{1}$ and $z_{2}$ split the system into two parts and in order to have $k_{x}\in\cof{n,n+1}$ $\forall x$, the number of shift-points must be even in both of these parts separately. Therefore, if we neglect the complications that arise at the boundaries between the two parts, the total weight of all configurations that can be obtained by adding shift-points to the maximum weight configuration in the left-hand panel, is approximately given by: $\sof{\sum_{m=0}^{\floor{\Delta z/2}}\,\binom{\Delta z/2}{2\,m}\,I_{0}^{\Delta z/2-2\,m}\of{\tilde{s}}\,I_{1}^{2\,m}\of{\tilde{s}}}\sof{\sum_{m=0}^{\floor{\of{N_{t}-\Delta z}/2}}\,\binom{\of{N_{t}-\Delta z}/2}{2\,m}\,I_{0}^{\of{N_{t}-\Delta z}/2-2\,m}\of{\tilde{s}}\,I_{1}^{2\,m}\of{\tilde{s}}}$, where $\Delta z=\of{z_{2}-z_{1}}$.} 
\label{fig:o2twopointpartf1dmonomerinsertion}
\end{figure*}

In order to better understand this behavior, we determine analytically the leading contributions to the one- and two-point functions at the critical points $\mu_{n,n+1}$, $n\in\mathbb{Z}$.\\
For $\mu=\mu_{n,n+1}$, the factors $I_{k_{x}}\of{\beta}\,\e^{2\,\mu\,k_{x}}$ in \eqref{eq:o2fluxreppartf1dincs} are also for $\tilde{s}>0$ maximized if $k_{x}\in\cof{n,n+1}$. If $\tilde{s}$ is not too large, we can then assume that the one- and two-point functions, \eqref{eq:o2cond1dfinites} and \eqref{eq:o2corr1dfinites}, are well approximated by taking into account only the 2 by 2 blocks of the transfer-matrix $\ssof{T_{\delta}}_{a b}$ from \eqref{eq:o2fluxrep1dtransfermat} for which $a,b\in\cof{n,n+1}$.\\
For the two-point function at time separation $t$ one then finds:
\begin{widetext}
\begin{align}
\savof{\phi^{-}_{0}\phi^{+}_{t}}&\of{\mu=\mu_{n,n+1},N_{t}}\nonumber\\
&\overset{\phantom{\text{\eqref{eq:binarymodelpartf}}}}{\approx}\,\frac{\sum\limits_{\cof{k_{i}\in\cof{n,n+1}}_{\forall i}}\,\sof{T_{-1}}_{k_{N_{t}-1},k_{0}}\sof{T_{0}}_{k_{0},k_{1}}\ldots\sof{T_{1}}_{k_{t-1},k_{t}}\sof{T_{0}}_{k_{t},k_{t+1}}\ldots}{\sum\limits_{\cof{k_{i}\in\cof{n,n+1}}_{\forall i}}\,\sof{T_{0}}_{k_{N_{t}-1},k_{0}}\ldots \sof{T_{0}}_{k_{N_{t}-2},k_{N_{t}-1}}}\nonumber\\
&\overset{\begin{subarray}{1}\text{\eqref{eq:o2fluxrep1dtransfermat}}\\
\text{\eqref{eq:o2critmu1d}}
\end{subarray}}{\approx}\,\frac{c\of{t,I_{0}\of{\tilde{s}},I_{1}\of{\tilde{s}}}\,c\of{N_{t}-t,I_{0}\of{\tilde{s}},I_{1}\of{\tilde{s}}}}{4\,c\of{N_{t},I_{0}\of{\tilde{s}},I_{1}\of{\tilde{s}}}}\nonumber\\
&\overset{\text{\eqref{eq:binarymodelpartf}}}{=}\,\frac{1}{8}\bof{1+\frac{\cosh\sof{\mathsmaller{\of{\frac{N_{t}}{2}-t}\log\of{\frac{I_{0}\of{\tilde{s}}-I_{1}\of{\tilde{s}}}{I_{0}\of{\tilde{s}}+I_{1}\of{\tilde{s}}}}}}}{\cosh\sof{\mathsmaller{\frac{N_{t}}{2}\underbrace{\mathsmaller{\mathsmaller{\log\of{\frac{I_{0}\of{\tilde{s}}-I_{1}\of{\tilde{s}}}{I_{0}\of{\tilde{s}}+I_{1}\of{\tilde{s}}}}}}}_{\mathclap{\substack{-\of{\log\ssof{I_{0}\of{\tilde{s}}+\totd{I_{0}\of{\tilde{s}}}{\tilde{s}}}-\log\ssof{I_{0}\of{\tilde{s}}-\totd{I_{0}\of{\tilde{s}}}{\tilde{s}}}}\\\approx -2\totd{\log\of{I_{0}\of{\tilde{s}}}}{\tilde{s}}=-\tilde{s}+\order{\tilde{s}^{3}}}}}}}}}\,
\approx\,\frac{1}{8}\bof{1+\frac{\cosh\sof{\of{N_{t}/2-t}\,\tilde{s}}}{\cosh\of{N_{t}\,\tilde{s}/2}}}\ ,\label{eq:o2corr1dfinitesapprox}
\end{align}
\end{widetext}
where the function $c\of{L,A,B}$ is given by
\begin{multline}
c\of{L,A,B}\,=\,\sum\limits_{m=0}^{\floor{L/2}}\,\binom{L}{2\,m}\,A^{L-2\,m}\,B^{2\,m}\\
=\,\frac{1}{2}\of{\of{A+B}^{L}\,+\,\of{A-B}^{L}}\ ,\label{eq:binarymodelpartf}
\end{multline}
and represents the partition function for a system of $L\in\mathbb{N}$ binary degrees of freedom, in which the two distinct states have weights $A$ and $B$, respectively, and where the states with weight $B$ must always be occupied by an even number of degrees of freedom. The reason for the appearance of the function $c\of{L,A,B}$ in \eqref{eq:o2corr1dfinitesapprox} is explained in Figs.~\ref{fig:o2partf1dmonomerinsertion}-\ref{fig:o2twopointpartf1dmonomerinsertion}; in short: to each site of our (0+1)-dimensional lattice corresponds one binary d.o.f. and the two possible states of the d.o.f. correspond to whether the two $k$-variables that touch at the site do ($\sim$state with weight $A$) or do not ($\sim$state with weight $B$) have the same value.\\

Along the same line, one finds the leading contributions to the one-point function at a critical point $\mu_{n,n+1}$ to be: 
\begin{widetext}
\begin{multline}
\savof{\phi^{\pm}}\of{\mu=\mu_{n,n+1},N_{t}}\,\approx\,\frac{\sum\limits_{\cof{k_{i}\in\cof{n,n+1}}_{\forall i}}\,\sof{T_{\pm 1}}_{k_{N_{t}-1},k_{0}}\sof{T_{0}}_{k_{0},k_{1}}\ldots\sof{T_{0}}_{k_{N_{t}-2},k_{N_{t}-1}}}{\sum\limits_{\cof{k_{i}\in\cof{n,n+1}}_{\forall i}}\,\sof{T_{0}}_{k_{N_{t}-1},k_{0}}\ldots \sof{T_{0}}_{k_{N_{t}-2},k_{N_{t}-1}}}\\
=\,\frac{c_{odd}\of{N_{t},I_{0}\of{\tilde{s}},I_{1}\of{\tilde{s}}}}{2^{3/2}\,c\of{N_{t},I_{0}\of{\tilde{s}},I_{1}\of{\tilde{s}}}}\,=\,-\frac{1}{2^{3/2}}\tanh\of{\mathsmaller{\frac{N_{t}}{2}\log\of{\frac{I_{0}\of{\tilde{s}}-I_{1}\of{\tilde{s}}}{I_{0}\of{\tilde{s}}+I_{1}\of{\tilde{s}}}}}}\\
\approx\,2^{-3/2}\tanh\of{N_{t}\,\tilde{s}/2}\ ,\label{eq:o2cond1dfinitesapprox}
\end{multline}
\end{widetext}
where $c\of{L,A,B}$ is again given by \eqref{eq:binarymodelpartf}, and the new function,
\begin{multline}
c_{odd}\of{L,A,B}\,=\\
\sum\limits_{m=0}^{\floor{\of{L-1}/2}}\,\binom{L}{2\,m+1}\,A^{L-\of{2\,m+1}}\,B^{2\,m+1}\\
=\,\frac{1}{2}\of{\of{A+B}^{L}\,-\,\of{A-B}^{L}}\ , \label{eq:binarymodelpartfodd}
\end{multline}
has a similar meaning as $c\of{L,A,B}$, but the state with weight $B$ has now to be occupied always by an odd instead of even number of degrees of freedom.\\


From \eqref{eq:o2corr1dfinitesapprox} and \eqref{eq:o2cond1dfinitesapprox} we can finally obtain an expression for the \emph{connected} two-point function at a critical point $\mu_{n,n+1}$:
\begin{multline}
\avof{\phi^{-}_{x}\phi^{+}_{y}}_{conn.}\of{\mu=\mu_{n,n+1},N_{t}}\\
=\of{\avof{\phi^{-}_{x}\phi^{+}_{y}}-\avof{\phi^{-}_{x}}\avof{\phi^{+}_{y}}}\of{\mu=\mu_{n,n+1},N_{t}}\\
\approx\frac{1}{8}\of{\frac{\cosh\sof{\of{N_{t}/2-t}\,\tilde{s}}}{\cosh\of{N_{t}\,\tilde{s}/2}}+1-\tanh^{2}\ssof{N_{t}\,\tilde{s}/2}}\\
\overset{N_{t}\tilde{s}\gg 1}{\approx}\frac{\cosh\sof{\of{N_{t}/2-t}\,\tilde{s}}}{8\,\cosh\of{N_{t}\,\tilde{s}/2}}\,\ ,\label{eq:o2fluxrepcritconncorr1dfinites}
\end{multline}
where on the last line we used that $\tanh^{2}\ssof{x}$ quickly approaches unity for $x>1$, so that the constant piece in $\ssavof{\phi^{-}_{x}\phi^{+}_{y}}$ gets cancelled. The reason why this cancellation happens only for sufficiently large $N_{t}\tilde{s}$ is, that the fields $\phi^{\pm}$ are, as mentioned earlier, not invariant under global $\Un{1}$-transformations and $\ssavof{\phi^{\pm}}$ therefore receives cancelling contributions form different $\Un{1}$-phases if $N_{t}\tilde{s}$ is not sufficiently large to give dominant weight to just a single phase. Only if the $\ssavof{\phi^{\pm}}$ are dominated by a single phase, the term $\ssavof{\phi^{+}}\ssavof{\phi^{-}}$ in \eqref{eq:o2fluxrepcritconncorr1dfinites} can fully cancel the disconnected diagrams in $\ssavof{\phi^{-}_{x}\phi^{+}_{y}}$.\\

Now let us return to the question of why the constant piece in the two-point function gets smaller when the source $\tilde{s}$ changes from zero to a non-zero value. The first thing to note is, that for $\tilde{s}>0$ the connected two-point function \eqref{eq:o2fluxrepcritconncorr1dfinites} at a critical value of $\mu$ does no-longer correspond to a zero-mode but to a mode of mass $\tilde{s}$. The cause of the constant pieces \eqref{eq:o2fluxrepcorrdisconnpiece1d} we observed at critical $\mu$ for $\tilde{s}=0$ is therefore gone. The constant pieces appearing in \eqref{eq:o2corr1dfinitesapprox} for $\tilde{s}>0$ are instead due to disconnected diagrams and can be removed by subtracting $\avof{\phi^{-}_{x}}\avof{\phi^{+}_{y}}$ in \eqref{eq:o2fluxrepcritconncorr1dfinites}.\\
By setting in \eqref{eq:o2corr1dfinitesapprox} the source $\tilde{s}$ to zero, the disconnected diagrams are gone and at the same time the massless modes reappear, so that we recover the result from \eqref{eq:o2fluxrepcorrdisconnpiece1d} for $\mu=\mu_{n,n+1}$. For $\tilde{s}>0$, on the other hand, the minimum in the two-point function at $t=N_{t}/2$ changes smoothly as a function of $N_{t}$ from the value $1/4$ for $N_{t}\ll \tilde{s}^{-1}$ to the value $1/8$ for $N_{t}\gg \tilde{s}^{-1}$.\\

From the derivation of the result in \eqref{eq:o2corr1dfinitesapprox}, it can be seen, that this decrease in the minimum of the two-point function as function of $N_{t}$ when $\tilde{s}>0$ (remember that $\tilde{s}$ still has to be small for our approximation with the restricted transfer-matrix to be valid), is due to configurations with disconnected diagrams, which start to contribute to both, the partition function as well as observables, but in different amounts. If we write $\ssavof{\phi^{-}_{x}\phi^{+}_{y}}=Z_{2}^{-,+}\of{x,y}/Z$ (cf. appendix~\ref{ssec:dualnpointfunc}), the number of configurations with disconnected diagrams, contributing to the partition function $Z$, is always larger than the number of configurations with disconnected diagrams, contributing to the two-point partition function $Z_{2}^{-,+}\of{x,y}$, as in the latter, the presence of the external fields at $x$ and $y$ puts additional constraints on the form of the possible disconnected diagrams. The ratio $Z_{2}^{-,+}\of{x,y}/Z$ has therefore to decrease when $\tilde{s}$ changes from zero to a non-zero value. For small $N_{t}$, the number of possible disconnected diagrams that can contribute to $Z$ or $Z_{2}^{-,+}\of{x,y}$, is relatively small, and because each discontinuity leads to a suppression factor $I_{1}\ssof{\tilde{s}}/I_{0}\ssof{\tilde{s}}$ in the weight of the corresponding configuration, disconnected configurations contribute then only marginally to the partition functions $Z$ and $Z_{2}^{-,+}\of{x,y}$. In this case, the fact, that there are less disconnected diagrams in $Z_{2}^{-,+}\of{x,y}$ than in $Z$ is irrelevant. With increasing $N_{t}$, the number of possible configurations with disconnected diagrams grows rapidly for both, $Z$ and $Z_{2}^{-,+}\of{x,y}$, while the number of fully connected configurations remains the same. For some sufficiently large $N_{t}$, the configurations with disconnected diagrams will therefore start to dominate in $Z$ and $Z_{2}^{-,+}\of{x,y}$ over the fully connected configurations, after which the fact, that the number of disconnected diagrams that can contribute to $Z$, is larger than the number of disconnected diagrams that can contribute to $Z_{2}^{-,+}\of{x,y}$, becomes relevant and at some point completely dominates the value of $Z_{2}^{-,+}\of{x,y}/Z$.\\

So far, we have for $\tilde{s}>0$ only discussed the $N_{t}$-dependency of condensate and correlation function in the case where $\beta$ is kept fixed. However, as the combination $\sigma:=\tilde{s}\,N_{t}\sim J/T$ is dimensionless (cf. text below eq.~\eqref{eq:discretederiv} for case $d=1$), the situation should remain qualitatively unchanged if we keep instead of $\beta$, the quantities $\kappa=\beta/N_{t}=T\,f_{\pi}^{2}$ and $\sigma=\tilde{s}\,N_{t}\sim J/T$ fixed while $N_{t}$ is increased, provided that $\kappa$ and $\sigma$ are sufficiently small, so that the approximations, that lead to the expressions \eqref{eq:o2corr1dfinitesapprox} and \eqref{eq:o2cond1dfinitesapprox} for two- and one-point function, respectively, remain valid.\\
For too large values of $\tilde{s}$, the coupling to the external field will dominate over the nearest-neighbour interactions and one expects that if the system develops long range order, the condensate should approach the following value:
\begin{multline}
\abs{\avof{\phi^{\pm}}}\,\approx\,\frac{1}{N_{t}\,\sqrt{2}}\partd{}{\tilde{s}}\log\bof{\int\limits_{-\pi}^{\pi}\dd\theta\,\e^{N_{t}\,\tilde{s}\cos\of{\theta-\theta_{s}}}}\\
=\,\frac{1}{\sqrt{2}}\frac{I_{1}\of{N_{t}\,\tilde{s}}}{I_{0}\of{N_{t}\,\tilde{s}}}\,\overset{\tilde{s} N_{t}\gg 1}{\approx}\,\frac{1}{\sqrt{2}}\,\approx\,0.7071\ .
\end{multline}
It can be numerically verified that the full expression \eqref{eq:o2cond1dfinites} reproduces this result, while our approximate expression \eqref{eq:o2cond1dfinitesapprox} clearly has to fail in doing so, as it can apparently not grow larger than $2^{-3/2}\approx 0.3536$. To improve the approximate formula \eqref{eq:o2cond1dfinitesapprox}, configurations that include $k$-variables with values that are different from $n$ or $n+1$, would have to be included, whose large number an variety would quickly become unmanageable in a manual computation. 

\section{Summary and discussion}\label{sec:discussion}
In Sec.~\ref{ssec:phasetransitions} we have seen, that the (0+1)-dimensional non-linear $\On{2}$-model at finite density undergoes phase-transitions between different vacua when the chemical potential $\mu$ crosses the critical values $\mu_{n,n+1}$, $n\in\mathbb{Z}$, given in \eqref{eq:o2critmu1d}. The different vacua form a discrete set and can be labeled by the integer-valued charge-density they carry. The two subscripts $n$ and $n+1$ in $\mu_{n,n+1}$ refer to these charge densities of the two vacua that are separated by the transition at $\mu=\mu_{n,n+1}$. \\

Sec.~\ref{ssec:corrfuncandmassspec} illustrated, that although the lattice system is one-dimensional (it extends only in time direction), the two-point function $\ssavof{\phi^{-}_{x}\phi^{+}_{y}}$ develops at the transition points $\mu=\mu_{n,n+1}$, $n\in\mathbb{Z}$ a non-zero constant piece of magnitude $1/4$, which is due to zero-modes and indicates the development of some sort of long-range order in the system.\\

In Sec.~\ref{ssec:explicitsymmbreak} we then discussed the case $\tilde{s}>0$: the non-zero source removes all zero-modes and instead allows configurations with disconnected diagrams to contribute to the partition function and to expectation values of observables. These disconnected diagrams replace the zero modes as cause of the constant piece in the two-point function when $\mu$ approaches a critical value $\mu_{n,n+1}$, $n\in\mathbb{Z}$, and in addition give rise to a non-zero condensate $\ssavof{\phi^{\pm}}$. For sufficiently large system sizes, both, the constant piece in the two-point function, as well as the condensate develop a pronounced peak when $\mu$ approaches one of the critical values $\mu_{n,n+1}$, $n\in\mathbb{Z}$.\\

These findings generalize to linear and non-linear $\On{N}$ spin models with $N\geq 2$; in most of the equations presented in Sec.~\ref{sec:results}, the Bessel functions $I_{k}\of{\beta}$ would just need to be replaced by more complicated link-weight functions.\\

As mentioned earlier, in the solid state physics interpretation of the partition function for our (0+1) dimensional $\On{2}$ model, we consider the Euclidean time-direction as spatial dimension and let, instead of $1/N_{t}$, the paramete $\beta$ play the role of inverse temperatuere: $\beta\propto 1/\ssof{k_{B}\,T}$. The Euclidean action then turns into $S=\beta\,H$, with $H$ being the Hamiltonian of a corresponding one-dimensional solid state physics $\On{2}$-spin system. In this context, the above mentioned findings of long range order and a non-zero condensate in this 1D $\On{2}$ spin system seem to conflict with the findings of Mermin-Wagner~\cite{Mermin} and Wegner~\cite{Wegner}, that an $\On{N}$-symmetric spin system with finite-range interactions can in one or two dimensions not undergo spontaneous magnetization, and not develop long-range order. Why does our model nevertheless show signs of long-range order at a discrete but infinite set of $\mu$-values?\\

We can come up with essentially three possible reasons why the Mermin-Wagner theorem does not need to apply in the present case:\\

The first reason is, that in the solid state physics interpretation given above, the lattice spacing can be considered as physical, and the thermodynamic limit is therefore always obtained by sending $N_{t}$ to infinity while the lattice spacing $a$ and the inverse temperature $\beta$ are kept fixed. But, as we have seen in Secs.~\ref{ssec:phasetransitions} and \ref{ssec:corrfuncandmassspec}, if $N_{t}$ is sent to infinity at fixed $\beta$, then the system undergoes first order phase transitions at the very same values of $\mu$ for which one can observe the appearance of long-range order.\\   

One could also argue, that the action of our model is complex for $\mu\neq 0$ when expressed in terms of the original spin variables, and it is not obvious, that the proof of the Mermin-Wagner theorem extends to this case. In the solid state physics interpretation of our $\On{2}$ model, the variables $\cof{\theta_{x}}_{x}$ can for $\mu=0$ be considered as physical degrees of freedom, which specify the configuration of a set of classical spins on a 1D lattice. Each configuration has in this case a well-defined energy and Boltzmann weight. But, as soon as $\mu\neq 0$, the Hamiltonian is in general complex and the $\theta$-variables loose their direct physical meaning, as configurations which are encoded in terms of physical degrees of freedom should take on only real energy values. One could therefore infer, that the system can escape the Mermin-Wagner theorem, because it is only the formulation in terms of these unphysical degrees of freedom that makes the Hamiltonian look as if the theorem should apply. But, if one parametrizes the system also for $\mu\neq 0$ in terms of \emph{physical} degrees of freedom, these physical degrees of freedom (and the ways in which they interact) do no-longer satisfy the prerequisites for the theorem to apply: in the flux-variables representation of the $\On{2}$ partition function \eqref{eq:o2fluxreppartfg} in (0+1) dimensions, the interaction between the $k$-variables that live on different links, becomes for example of infinite range when $\tilde{s}=0$, due to the conservation of the $\Un{1}$-flux, which renders the Mermin-Wagner theorem inapplicable~\cite{Thouless,Imry}.\\

Finally, a third argument for why the Mermin-Wagner theorem does not need to apply to our finite density $\On{2}$ model, is motivated by the fact that the charge density in our $\On{2}$ system would in continuous Minkowski space be given by the time-component of the conserved current $j^{\nu}\propto \ii\,\sof{\ssof{\partial^{\nu}\phi^{-}\of{x}}\phi^{+}\of{x}-\ssof{\partial^{\nu}\phi^{+}\of{x}}\phi^{-}\of{x}}=\partial^{\nu}\theta\of{x}$ (up to a divergence-free vector field, which in (0+1) dimensions is just a constant). One might therefore assume that the different vacua of the corresponding Euclidean lattice model, which carry different integer-valued charge densities, differ by how often $\theta_{x}$ wraps around the interval $\loint{-\pi,\pi}$ as $x$ moves along a path that winds in time direction around the periodic lattice.\\
This \emph{per lattice winding}, $N_{\Omega}$, is the Euclidean time integral of the \emph{charge density} $n$, and one can therefore write: $n=N_{\Omega}/N_{t}$, where $N_{t}$ is the temporal extent of the periodic lattice. An integer charge density $n$ therefore implies that the $\theta_{x}$ variable had to wrap $n\cdot N_{t}$ times around the interval $\loint{-\pi,\pi}$ as $x$ wraps once around the periodic lattice in time-direction. This means that $\theta_{x}$ should undergo $n$ full windings while traversing a single link. Such a \emph{per link winding} can, of course, not be represented by the lattice field $\theta_{x}$, as $\Delta\theta_{x,d}=\ssof{\theta_{x+\hat{d}}-\theta_{x}}$ has to satisfy $\ssabs{\Delta\theta_{x,d}}<2\,\pi$ if $\theta_{x}\in\loint{-\pi,\pi}$ $\forall x$. But, we can for the moment consider the values $\theta_{x}$ and $\theta_{x+\hat{d}}$ of the lattice field on adjacent sites of a link $\ssof{x,d}$ that connects the sites $x$ and $x+\hat{d}$ as boundary conditions for a 1d continuum field $\theta\of{x}$ that lives on the link $\ssof{x,d}$ and interpolates between the two boundary values. The number of windings that this continuum field undergoes along the link $\ssof{x,d}$, will then have an impact on how strongly the action for this continuum field changes when $\Delta\theta_{x,d}$ is changed. So, a change in the winding number of the continuum field on the link $\ssof{x,d}$ would imply a change in the interaction strength between the lattice variables $\theta_{x}$ and $\theta_{x+\hat{d}}$. As the continuum field will try to behave classically, the winding number will grow proportionally to the chemical potential (up to the fact that it can only change in discrete steps) and one can therefore imagine that in the lattice formulation, the per link winding number of the link $\ssof{x,d}$ is encoded in the combined values of $\mu$ and $\Delta\theta_{x,d}$.\\
This suggests that the long-range order we observe in our (0+1)-dimensional $\On{2}$ lattice system whenever $\mu$ approaches one of the critical values $\mu_{n,n+1}$ from \eqref{eq:o2critmu1d}, with $n\in\mathbb{Z}$, is not of the usual form, i.e. not related to the spontaneous breaking of a global symmetry and the formation of a homogeneous non-zero vacuum expectation value, but rather occurs because the magnitude of the chemical potential is such that the corresponding preferred phase velocity $\dot{\theta}\of{x}$ of the $\phi^{\pm}$-fields is so high, that $\phi\of{x}$ undergoes multiple windings as $x$ moves along the distance of a single lattice spacing. In the field-theory context, this is clearly a lattice artefact, which occurs because the energy scale set by $\mu$ exceeds the lattice cut-off scale. In the above mentioned solid state physics context, however, where the one-dimensional lattice is spatial and the chemical potential $\mu$ might play the role of a negative resistance, the lattice spacing is physical and does not define an energy cut-off. The magnitude of $\mu$ is therefore in this case not restricted.  

\section{Acknowledgement}
I would like to thank Philippe de Forcrand for providing helpful comments on an earlier version of this manuscript, which helped improving clarity.

\begin{appendices}
\renewcommand{\thesubsection}{\arabic{subsection}}
\section{Flux-variable formulation of $\On{N}$ spin model}\label{sec:fluxvardualization}
In this appendix, we review, following \cite{Bruckmann,Rindlisbacher1,Rindlisbacher3}, how the flux variable representation for the non-linear $\On{N}$ spin model with a chemical potential and source terms is obtained.

\subsection{Dualization of the partition function}\label{ssec:dualizationofpartf}
Starting point is the partition function for the action \eqref{eq:onaction}:
\begin{widetext}
\[
Z\,=\,\int\DD{\phi}\,\exp\of{\sum\limits_{x}\bcof{\frac{\beta}{2}\,\sum\limits_{\nu=1}^{d}\of{\phi_{x}\,\e^{2\,\mu\,\tau_{1 2}\,\delta_{\nu,d}}\,\phi_{x+\hat{\nu}}+\phi_{x}\,\e^{-2\,\mu\,\tau_{1 2}\,\delta_{\nu,d}}\,\phi_{x-\hat{\nu}}}+\of{s\cdot\phi_{x}}}}\ ,\label{eq:onpartf}
\]
\end{widetext}
with $\phi_{x}\in\Sph{N-1}\subset\mathbb{R}^{N}$. To carry out the dualization, we now write the exponential in \eqref{eq:onpartf} as a product of separate exponential factors for each individual term in the action:
\begin{widetext}
\begin{multline}
Z\,=\,\int\DD{\phi}\prod\limits_{x}\bcof{\bof{\prod\limits_{\nu=1}^{d}\exp\sof{\beta\,\e^{2\,\mu\,\delta_{\nu,d}}\,\phi^{-}_{x}\,\phi^{+}_{x+\hat{\nu}}}\exp\sof{\beta\,\e^{-2\,\mu\,\delta_{\nu,d}}\,\phi^{-}_{x+\hat{\nu}}\,\phi^{+}_{x}}\\
\cdot\bof{\prod\limits_{i=3}^{N}\,\exp\sof{\beta\,\phi^{i}_{x}\,\phi^{i}_{x+\hat{\nu}}}}}\exp\sof{s^{+}\,\phi^{+}_{x}}\exp\sof{s^{-}\,\phi^{-}_{x}}\,\bof{\prod\limits_{i=3}^{N}\exp\sof{s^{i}\,\phi^{i}_{x}}}}\ ,\label{eq:phifourfluxrepd1}
\end{multline}
\end{widetext}
where $\phi^{\pm}=\frac{1}{\sqrt{2}}\of{\phi_{1}\pm\ii\,\phi_{2}}$ and $s^{\pm}=\frac{1}{\sqrt{2}}\of{s_{1}\mp\ii\,s_{2}}$. Then we write these exponentials in terms of power-series:
\begin{subequations}\label{eq:onexpexpand}
\begin{multline}
\exp\sof{\beta\,\e^{2\,\mu\delta_{\nu,d}}\,\phi^{-}_{x}\,\phi^{+}_{x+\hat{\nu}}}\,=\\
\sum\limits_{\eta_{x,\nu}}\frac{\sof{\phi^{-}_{x}\,\phi^{+}_{x+\hat{\nu}}\,\beta\,\e^{2\,\mu\delta_{\nu,d}}}^{\eta_{x,\nu}}}{\eta_{x,\nu}!}\ ,
\end{multline}
\begin{multline}
\exp\sof{\beta\,\e^{-2\,\mu\delta_{\nu,d}}\,\phi^{+}_{x}\,\phi^{-}_{x+\hat{\nu}}}\,=\\
\sum\limits_{\bar{\eta}_{x,\nu}}\frac{\sof{\phi^{+}_{x}\,\phi^{-}_{x+\hat{\nu}}\,\beta\,\e^{-2\,\mu\delta_{\nu,d}}}^{\bar{\eta}_{x,\nu}}}{\bar{\eta}_{x,\nu}!}\ ,
\end{multline}
\[
\exp\sof{\beta\,\phi^{i}_{x}\,\phi^{i}_{x+\hat{\nu}}}=\sum\limits_{\chi^{i}_{x,\nu}}\frac{\sof{\beta\,\phi^{i}_{x}\,\phi^{i}_{x+\hat{\nu}}}^{\chi^{i}_{x,\nu}}}{\chi^{i}_{x,\nu}!}\ ,
\]
and
\[
\exp\of{s^{+}\,\phi^{+}_{x}}\,=\,\sum\limits_{m_{x}}\frac{\sof{s^{+}\,\phi^{+}_{x}}^{m_{x}}}{m_{x}!}\ ,
\]
\[
\exp\of{s^{-}\,\phi^{-}_{x}}\,=\,\sum\limits_{\bar{m}_{x}}\frac{\sof{s^{-}\,\phi^{-}_{x}}^{\bar{m}_{x}}}{\bar{m}_{x}!}\ ,
\]
\[
\exp\of{s^{i}\,\phi^{i}_{x}}\,=\,\sum\limits_{n^{i}_{x}}\frac{\sof{s^{i}\,\phi^{i}_{x}}^{n^{i}_{x}}}{n^{i}_{x}!}\ ,
\]
\end{subequations}
and after writing the $\phi_{x}$ in polar form, i.e.
\begin{subequations}
\begin{align}
\phi^{+}_{x}\,&=\,\sin\sof{\theta^{\of{N-1}}_{x}}\,\cdots\,\sin\sof{\theta^{\of{2}}_{x}}\,\frac{\e^{\ii\,\theta^{\of{1}}_{x}}}{\sqrt{2}}\\
\phi^{-}_{x}\,&=\,\sin\sof{\theta^{\of{N-1}}_{x}}\,\cdots\,\sin\sof{\theta^{\of{2}}_{x}}\,\frac{\e^{-\ii\,\theta^{\of{1}}_{x}}}{\sqrt{2}}\\
\phi^{3}_{x}\,&=\,\sin\sof{\theta^{\of{N-1}}_{x}}\,\cdots\,\cos\sof{\theta^{\of{2}}_{x}}\\
 &\hspace{5pt}\vdots \nonumber\\
\phi^{N-1}_{x}\,&=\,\sin\sof{\theta^{\of{N-1}}_{x}}\,\cos\sof{\theta^{\of{N-2}}_{x}}\\
\phi^{N}_{x}\,&=\,\cos\sof{\theta^{\of{N-1}}_{x}}\ ,
\end{align}
\end{subequations}
and using the corresponding integration measure,
\begin{multline}
\DD{\phi}\propto\\
\prod\limits_{x}\,\bof{\prod\limits_{i=2}^{N-1}\,\sof{\sin\sof{\theta_{x}^{\of{i}}}}^{i-1}}\,\dd{\theta_{x}^{\of{1}}}\wedge\ldots\wedge \dd{\theta_{x}^{\of{N-1}}}\ ,
\end{multline}
the partition function \eqref{eq:onpartf} can be written as
\begin{widetext}
\begin{multline}
Z\,=\,\sum\limits_{\cof{\eta,\bar{\eta},m,\bar{m}}}\prod\limits_{x}\bcof{\bof{\prod\limits_{\nu}\frac{\sof{\frac{\beta}{2}}^{\eta_{x,\nu}+\bar{\eta}_{x,\nu}}}{\eta_{x,\nu}!\,\bar{\eta}_{x,\nu}!}}\frac{\sof{\frac{s^{+}}{\sqrt{2}}}^{m_{x}}\sof{\frac{s^{-}}{\sqrt{2}}}^{\bar{m}_{x}}}{m_{x}!\,\bar{m}_{x}!}\,\bof{\prod\limits_{i=3}^{N}\bof{\prod\limits_{\nu}\frac{\beta^{\chi^{i}_{x,\nu}}}{\chi^{i}_{x,\nu}!}}\frac{\sof{s^{i}}^{n^{i}_{x}}}{n^{i}_{x}!}}\\
\cdot\e^{2\,\mu\of{\eta_{x,d}-\bar{\eta}_{x,d}}}\,\int\limits_{-\pi}^{\pi}\idd{\theta_{x}^{\of{1}}}{}\,\e^{\ii\,\theta^{\of{1}}_{x}\,\sof{m_{x}-\bar{m}_{x}-\sum\limits_{\nu}\of{\eta_{x,\nu}-\bar{\eta}_{x,\nu}-\of{\eta_{x-\hat{\nu},\nu}-\bar{\eta}_{x-\hat{\nu},\nu}}}}}\\
\cdot\bof{\prod\limits_{j=2}^{N-1}\,\int\limits_{0}^{\pi}\idd{\theta_{x}^{\of{j}}}{}\,\sof{\cos\sof{\theta_{x}^{\of{j}}}}^{\sum\limits_{\nu}\sof{\chi^{j+1}_{x,\nu}+\chi^{j+1}_{x-\hat{\nu},\nu}}}\\
\cdot\sof{\sin\sof{\theta_{x}^{\of{j}}}}^{j-1+\sum\limits_{\nu}\sof{\eta_{x,\nu}+\bar{\eta}_{x,\nu}+\eta_{x-\hat{\nu},\nu}+\bar{\eta}_{x-\hat{\nu},\nu}+\sum\limits_{i=3}^{j}\sof{\chi^{i}_{x,\nu}+\chi^{i}_{x-\hat{\nu},\nu}}}}}}\ .\label{eq:onfluxrepd2}
\end{multline}
\end{widetext}
To simplify the notation, we define new variables $k_{x,\nu}\in \mathbb{Z}$ and $l_{x,\nu}\in \mathbb{N}_{0}$, so that:
\begin{subequations}\label{eq:klvariables2}
\begin{align}
\eta_{x,\nu}-\bar{\eta}_{x,\nu}&=\,k_{x,\nu}\ ,\\
\eta_{x,\nu}+\bar{\eta}_{x,\nu}&=\,\abs{k_{x,\nu}}+2\,l_{x,\nu}\ ,
\end{align}
\end{subequations}
and $p_{x}\in \mathbb{Z}$ and $q_{x}\in \mathbb{N}_{0}$, so that:
\begin{subequations}\label{eq:pqvariables2}
\begin{align}
m_{x}-\bar{m}_{x}&=\,p_{x}\ ,\\
m_{x}+\bar{m}_{x}&=\,\abs{p_{x}}+2\,q_{x}\ ,
\end{align}
\end{subequations}
as well as on each site $x$ the quantities
\[
A_{x}=\sum\limits_{\nu}\of{\sabs{k_{x,\nu}}+\sabs{k_{x-\hat{\nu},\nu}}+2\sof{l_{x,\nu}+l_{x-\hat{\nu},\nu}}}
\]
and
\[
B^{i}_{x}\,=\,\sum\limits_{\nu}\of{\chi^{i}_{x,\nu}+\chi^{i}_{x-\hat{\nu},\nu}}\ .
\]
Now we carry out the angular integrals in \eqref{eq:onfluxrepd2}, using that for $M,N\,\in\,\mathbb{N}_{0}$, we have\footnote{Compare \eqref{eq:genbetafunc} with the integral form of Euler's beta function, $B\of{m,n}=\frac{\Gamma\of{m}\Gamma\of{n}}{\Gamma\of{m+n}}=2\,\int_{0}^{\pi/2}\dd{\theta}\cos^{2\,m-1}\of{\theta}\,\sin^{2\,n-1}\of{\theta}$, and use the symmetry properties of the trigonometric functions with respect to reflection at $\pi/2$.}:
\begin{multline}
\int\limits_{0}^{\pi}\idd{\theta}{}\sin^{M}\of{\theta}\,\cos^{N}\of{\theta}\,=\\
\frac{1+\of{-1}^{N}}{2}\frac{\Gamma\sof{\frac{1+M}{2}}\,\Gamma\sof{\frac{1+N}{2}}}{\Gamma\sof{\frac{2+M+N}{2}}}\ ,\label{eq:genbetafunc}
\end{multline}
which yields:
\begin{widetext}
\begin{subequations}\label{eq:onfluxreppartf}
\begin{multline}
Z\,=\,\sum\limits_{\cof{k,l,\chi,p,q,n}}\prod\limits_{x}\bcof{\bof{\prod\limits_{\nu=1}^{d}\frac{\beta^{\abs{k_{x,\nu}}+2\,l_{x,\nu}+\sum_{i=3}^{N}\,\chi^{i}_{x,\nu}}}{\of{\abs{k_{x,\nu}}+l_{x,\nu}}!\,l_{x,\nu}!\,\prod_{i=3}^{N}\chi^{i}_{x,\nu}!}}\\
\cdot\frac{\ssof{s^{+}}^{\frac{1}{2}\of{\abs{p_{x}}+p_{x}}+q_{x}}\ssof{s^{-}}^{\frac{1}{2}\of{\abs{p_{x}}-p_{x}}+q_{x}}\,\e^{2\,\mu\,k_{x,d}}}{\of{\abs{p_{x}}+q_{x}}!\,q_{x}!}\bof{\prod\limits_{i=3}^{N}\frac{\ssof{s^{i}}^{n^{i}_{x}}}{n^{i}_{x}!}}\,\delta\sof{p_{x}-\sum\limits_{\nu}\sof{k_{x,\nu}-k_{x-\hat{\nu},\nu}}}\\
\cdot W\sof{A_{x}+\abs{p_{x}}+2\,q_{x},\,B^{3}_{x}+n^{3}_{x},\,\ldots,\,B^{N}_{x}+n^{N}_{x}}}\ ,
\end{multline}
or, with $s^{\pm}=\frac{\tilde{s}\e^{\mp\ii\phi_{s}}}{\sqrt{2}}$: 
\begin{multline}
Z\,=\,\sum\limits_{\cof{k,l,\chi,p,q,n}}\prod\limits_{x}\bcof{\bof{\prod\limits_{\nu=1}^{d}\frac{\beta^{\abs{k_{x,\nu}}+2\,l_{x,\nu}+\sum_{i=3}^{N}\,\chi^{i}_{x,\nu}}}{\of{\abs{k_{x,\nu}}+l_{x,\nu}}!\,l_{x,\nu}!\,\prod_{i=3}^{N}\chi^{i}_{x,\nu}!}}\\
\cdot\frac{\tilde{s}^{\abs{p_{x}}+2\,q_{x}}\,\e^{\ii\,\phi_{s}\,p_{x}}\,\e^{2\,\mu\,k_{x,d}}}{2^{\sfrac{\ssof{\abs{p_{x}}+2\,q_{x}}}{2}}\of{\abs{p_{x}}+q_{x}}!\,q_{x}!}\bof{\prod\limits_{i=3}^{N}\frac{\ssof{s^{i}}^{n^{i}_{x}}}{n^{i}_{x}!}}\,\delta\sof{p_{x}-\sum\limits_{\nu}\sof{k_{x,\nu}-k_{x-\hat{\nu},\nu}}}\\
\cdot W\sof{A_{x}+\abs{p_{x}}+2\,q_{x},\,B^{3}_{x}+n^{3}_{x},\,\ldots,\,B^{N}_{x}+n^{N}_{x}}}\ .
\end{multline}
\end{subequations}
where
\[
\delta\of{x}=\ucases{1\quad\text{if}\quad x=0\\0\quad\text{else}}\quad\text{and}\quad W\sof{A,B^{3},\ldots,B^{N}}\,=\,\frac{\Gamma\sof{\frac{2+A}{2}}\,\prod\limits_{i=3}^{N}\frac{1+\of{-1}^{B^{i}}}{2}\Gamma\sof{\frac{1+\vphantom{B}\smash{B^{i}}}{2}}}{2^{\sfrac{A}{2}}\,\Gamma\sof{\frac{N+A+\sum_{i=3}^{N}B^{i}}{2}}}\ .\label{eq:onweightfunc0}
\]
\end{widetext}
Not that the Taylor expansions of the exponential functions in \eqref{eq:onexpexpand} always converge, so that the two expressions \eqref{eq:onfluxreppartf} are exact rewritings of the original lattice partition function \eqref{eq:onpartf}. 

\subsection{One- and two-point functions in terms of flux-variables}\label{ssec:dualnpointfunc}
The one- and two-point functions (and in principle any other n-point function) can be obtained by promoting the sources $s^{i}$, $i\in\cof{+,-,3,\ldots,N}$ to lattice fields $s^{i}_{x}$ and taking derivatives of the logarithm of the partition function \eqref{eq:onfluxreppartf} with respect to the $s^{i}_{x}$ on different sites $x$:
\[
\avof{\phi^{i}_{x}}\,=\,\partd{\log\of{Z}}{s_{x}^{i}}\,=\,\frac{Z_{1}^{i}\of{x}}{Z}\ ,
\]
\begin{multline}
\avof{\phi^{i}_{x}\phi^{j}_{y}}-\avof{\phi^{i}_{x}}\avof{\phi^{j}_{y}}\,=\,\spartd{\log\of{Z}}{s^{i}_{x}}{s^{j}_{y}}\\
=\,\frac{Z_{2}^{i j}\of{x,y}}{Z}-\frac{Z_{1}^{i}\of{x}}{Z}\frac{Z_{1}^{j}\of{y}}{Z}\ ,
\end{multline}
where the expressions for the \emph{one- and two-point partition functions}, 
\[
Z^{i}_{1}\of{z}=\partd{Z}{s^{i}_{z}}\quad\text{and}\quad Z^{i j}_{2}\of{z_{1},z_{2}}=\spartd{Z}{s^{i}_{z_{1}}}{s^{j}_{z_{2}}}\ ,
\]
in terms of the flux- and monomer-variables are most easily obtained, by applying the derivatives to the original partition function \eqref{eq:onpartf} and then go again through the dualization steps described in the previous section (Sec.~\ref{ssec:dualizationofpartf}). Each derivative of \eqref{eq:onpartf} with respect to a source $s^{i}_{z}$ brings down an extra factor of $\phi^{i}_{z}$, which, in the course of the dualization procedure translates into integer shifts in the arguments of the site weights given by \eqref{eq:onweightfunc0} and the delta function. For the \emph{one-point partition function} one then finds:
\begin{widetext}
\begin{multline}
Z^{i}_{1}\of{z}\,=\,\sum\limits_{\cof{k,l,\chi,p,q,n}}\prod\limits_{x}\bcof{\bof{\prod\limits_{\nu=1}^{d}\frac{\beta^{\abs{k_{x,\nu}}+2\,l_{x,\nu}+\sum_{i=3}^{N}\,\chi^{i}_{x,\nu}}}{\of{\abs{k_{x,\nu}}+l_{x,\nu}}!\,l_{x,\nu}!\,\prod_{i=3}^{N}\chi^{i}_{x,\nu}!}}\\
\cdot\frac{\ssof{s^{+}}^{\frac{1}{2}\of{\abs{p_{x}}+p_{x}}+q_{x}}\ssof{s^{-}}^{\frac{1}{2}\of{\abs{p_{x}}-p_{x}}+q_{x}}\,\e^{2\,\mu\,k_{x,d}}}{\of{\abs{p_{x}}+q_{x}}!\,q_{x}!}\bof{\prod\limits_{i=3}^{N}\frac{\ssof{s^{i}}^{n^{i}_{x}}}{n^{i}_{x}!}}\\
\cdot\delta\sof{p_{x}{\color{fhl}-\of{\delta^{i,+}-\delta^{i,-}}\delta_{x,z}}-\sum\limits_{\nu}\sof{k_{x,\nu}-k_{x-\hat{\nu},\nu}}}\\
\cdot W\sof{A_{x}+\abs{p_{x}}+2\,q_{x}{\color{fhl}+\of{\delta^{i,+}+\delta^{i,-}}\delta_{x,z}},\,B^{3}_{x}+n^{3}_{x}{\color{fhl}+\delta^{i,3}\delta_{x,z}},\,\ldots,\,B^{N}_{x}+n^{N}_{x}{\color{fhl}+\delta^{i,N}\delta_{x,z}}}}\ ,\label{eq:onepointpartf}
\end{multline}
\end{widetext}
and similarly, for the \emph{two-point partition function}:
\begin{widetext} 
\begin{multline}
Z^{i j}_{2}\of{z_{1},z_{2}}\,=\,\sum\limits_{\cof{k,l,\chi,p,q,n}}\prod\limits_{x}\bcof{\bof{\prod\limits_{\nu=1}^{d}\frac{\beta^{\abs{k_{x,\nu}}+2\,l_{x,\nu}+\sum_{i=3}^{N}\,\chi^{i}_{x,\nu}}}{\of{\abs{k_{x,\nu}}+l_{x,\nu}}!\,l_{x,\nu}!\,\prod_{i=3}^{N}\chi^{i}_{x,\nu}!}}\\
\cdot\frac{\ssof{s^{+}}^{\frac{1}{2}\of{\abs{p_{x}}+p_{x}}+q_{x}}\ssof{s^{-}}^{\frac{1}{2}\of{\abs{p_{x}}-p_{x}}+q_{x}}\,\e^{2\,\mu\,k_{x,d}}}{\of{\abs{p_{x}}+q_{x}}!\,q_{x}!}\bof{\prod\limits_{i=3}^{N}\frac{\ssof{s^{i}}^{n^{i}_{x}}}{n^{i}_{x}!}}\\
\cdot\delta\sof{p_{x}{\color{fhl}-\of{\delta^{i,+}-\delta^{i,-}}\delta_{x,z_{1}}-\of{\delta^{j,+}-\delta^{j,-}}\delta_{x,z_{2}}}-\sum\limits_{\nu}\sof{k_{x,\nu}-k_{x-\hat{\nu},\nu}}}\\
\cdot W\sof{A_{x}+\abs{p_{x}}+2\,q_{x}{\color{fhl}+\of{\delta^{i,+}+\delta^{i,-}}\delta_{x,z_{1}}+\of{\delta^{j,+}+\delta^{j,-}}\delta_{x,z_{2}}}\,,\qquad\\
\qquad B^{3}_{x}+n^{3}_{x}{\color{fhl}+\delta^{i,3}\delta_{x,z_{1}}+\delta^{j,3}\delta_{x,z_{2}}},\,\ldots,\,B^{N}_{x}+n^{N}_{x}{\color{fhl}+\delta^{i,N}\delta_{x,z_{1}}+\delta^{j,N}\delta_{x,z_{2}}}}}\ .\label{eq:twopointpartf}
\end{multline}
\end{widetext}
The products of Kronecker deltas that are inserted in \eqref{eq:onepointpartf} and \eqref{eq:twopointpartf}, can be interpreted as \emph{external fields} inserted into the system at particular positions. The $Z^{i}_{1}\of{z}$ and $Z^{i j}_{2}\of{z_{1},z_{2}}$ represent therefore partition functions for the system defined by $Z$, but in the presence of external fields; receiving contributions only from configurations which are compatible with the presence of the inserted external fields.\\
In a similar way one can define arbitrary $n$-point partition functions $Z_{n}^{i_{1}\cdots i_{n}}\of{z_{1},\ldots,z_{n}}$. But, it should be noted, that the one-point partition functions \eqref{eq:onepointpartf} can only take on non-zero values if their corresponding sources $s^{i}$ are non-zero. This is due to the fact, that the world lines, which start or end at the inserted external field, also have to end somewhere, which is possible only if the system is able to produce an appropriate monomer (which is the case only if the corresponding source $s^{i}$ is non-zero). This is just a manifestation of the fact that in order to have a non-zero condensate, the vacuum needs to have overlap with the field-operator. For the same reason also all other $n$-point functions can be non-zero only if the inserted external fields $\phi^{i}_{z_{i}}$ come in pairs of particle and corresponding anti-particle (for $i\in\cof{\pm}$, this means $\phi^{\pm}_{z_{1}}\phi^{\mp}_{z_{2}}$-pairs), or if the corresponding source $s^{i}$ is non-zero. 
\end{appendices}

\end{document}